\documentclass[9pt,reprint, amsmath,amssymb, aps, twocolumn]{revtex4-2}

\usepackage{graphicx}
\usepackage{dcolumn}
\usepackage{bm}

\usepackage{amsmath,amsfonts,amssymb,amsthm}
\usepackage{tikz}
\usetikzlibrary{quantikz2}
\usepackage{graphicx,textcomp}
\usepackage{algorithm,algpseudocode}
\usepackage[colorlinks=true, allcolors=blue]{hyperref}
\usepackage{multirow}
\usepackage{arydshln}
\usepackage{caption}

\usepackage{soul}
\usepackage{xcolor}

\newcommand{\tr}{\textup{tr}}

\DeclareRobustCommand{\stirling}{\genfrac\{\}{0pt}{}}

\definecolor{lightyellow}{rgb}{1 1 .5}
\sethlcolor{lightyellow}

\begin{document}

\preprint{APS/123-QED}

\title{Heralded quasi-deterministic entanglement sources based on spontaneous parametric down-conversion}

\author{Yousef K. Chahine${}^{1}$}
\author{J. Gabriel Richardson${}^{2}$}
\author{Evan J. Katz${}^{1}$}%
\author{Adam J. Fallon${}^{1}$}%
\author{John D. Lekki${}^{1}$}%
\affiliation{${}^{1}$NASA Glenn Research Center, Cleveland, OH}
\affiliation{${}^{2}$University of Maryland, College Park, MD}

\begin{abstract}
A double-heralding technique is presented for producing heralded entangled photon pairs from spontaneous parametric down-conversion (SPDC).  Compared to the swap-heralded schemes studied in previous cascaded SPDC and zero-added-loss multiplexing (ZALM) proposals, this double-heralding technique is found to yield the most resource-efficient implementation in terms of minimizing the total number of sources and detectors required to achieve a specified rate and fidelity.  This is achieved by reducing the number of modes and mode-sorting optics needed on the heralding path.  Specifically, by immediately detecting any two signal photons from an array of down-converters, the corresponding idler photons can be projected onto an anti-correlated pair state which is shown to be unitarily equivalent to the state produced by swap-heralded sources, and hence can be used directly for long-range entanglement distribution in a ZALM architecture.  Quasi-deterministic operation through two distinct multiplexing techniques is analyzed.  The analysis derives expressions for the heralded pair probability and fidelity assuming realistic detectors with losses, dark counts, and partial photon number resolution (PNR), providing a framework for implementation of the source on a photonic integrated circuit (PIC).
\end{abstract}


\maketitle

\section{Introduction}

The development of reliable, high-rate sources of entangled photon pairs remains an area of fundamental research towards realizing many applications of quantum information science including linear optical quantum computing \cite{KNILL2001,BARTOLUCCI2023,BARTOLUCCI2021} as well as distributed quantum computing and quantum sensing with quantum networks \cite{KOMAR2014,KHABIBOULLINE2019}.  In particular, sources based on spontaneous parametric down-conversion (SPDC) and four-wave mixing (SFWM) remain strong candidates for producing entangled pairs at the rates necessary to support a useful quantum network.  However, SPDC sources have limitations as entangled pair sources arising from their probabilistic nature; namely, the noise generated by multi-pair emissions requires that the pair generation probability be made small, and thus high-fidelity, low-noise operation comes at the cost of a large effective loss from the source.

One method that has recently been investigated to improve SPDC for entanglement distribution is to use two polarization-entangled SPDC sources combined with a partial Bell state measurement (BSM), effectively using a local entanglement swap to herald the production of an entangled pair \cite{DHARA2022}.  This is analogous to the standard procedure for leveraging SPDC as a heralded single-photon source.  The heralding signal from a swap-heralded source ensures the presence of exactly one pair, enabling near-deterministic behavior through multiplexing \cite{SCOTT2020}.  This allows more efficient use of limited network resources, such as quantum memories loaded by the receivers in the ZALM scheme introduced by Chen et al. \cite{CHEN2023}.  The ZALM architecture takes advantange of SPDC sources with a large phase-matching bandwidth by channelizing the broadband down-converted photons using wavelength division multiplexing.  The scheme yields a high probability of producing a pair from a single pump pulse despite maintaining the low mean photon number per mode needed to limit noise from higher order emissions \cite{SHAPIRO2024}.  A frequency-resolved partial BSM between two such channelized photonic states allows the swap-heralded pair to be mode-matched to quantum memory via frequency-conversion and bandwidth-compression techniques.

In this work, we develop two generalizations of the aforementioned swap-heralded SPDC sources.  First, we introduce an alternative double-heralding scheme which eschews the need for an entanglement swap between polarization-entangled sources, instead relying on a simple double-detection event in any two signal modes of an array of independent SPDC processes.  This double-heralded approach moves the path-erasure process from the heralding modes to the heralded modes, thereby reducing the mode-sorting optics needed on the heralding path and maximizing the heralding efficiency which is critical to the fidelity of the source.  Second, we generalize the manner in which single photons can be heralded from two-mode squeezed vacuum (TMSV) sources, introducing the concept of heralded multimode TMSV ($M$-TMSV), with unique pair generation statistics depending on the number of modes $M$.  It is shown that swap-heralded sources realize the $M=2$ case.  A complete analysis of double-heralded $M$-TMSV sources yields closed form expressions for the heralding probability and heralded pair fidelity, including a treatment of all higher order emissions, losses, dark counts, and partial photon number resolution.

The double-heralding scheme quadratically reduces the number of multiplexed sources needed to produce a pair compared to the original ZALM and cascaded SPDC proposals \cite{DHARA2022,CHEN2023}.  This reduction is based on the observation that the heralded photons from two sources can be combined \emph{after} the double-heralding signal specifies which sources have produced a pair. This quadratic reduction of resources has also recently been developed by Shapiro \emph{et. al.} for the swap-heralded configuration \cite{SHAPIRO2025,EMBLETON2025}, and is referred to as cross-island heralding in the context of the frequency-multiplexed realization of the source.

After reviewing the swap-heralded pair source in Sec. \ref{sec:swapheralded}, the double-heralded TMSV scheme is introduced in Sec. \ref{sec:doubleheralded}.  Subtle differences in the quantum state produced by the two protocols leads us to the generalization to double-heralded $M$-TMSV in Sec. \ref{sec:doubleheraldedmtmsv}.  Analysis of the maximum multiplexed pair generation rate of heralded $M$-TMSV without heralding losses is presented in Sec. \ref{sec:doubleheralding}.  With realistic detectors the pair emission probability of the source must be tuned together with the characteristics of the heralding detectors to handle both the higher-order photon emissions and lower bounds imposed by dark counts.  The former limits the pair generation rate and the latter puts an upper bound on the pair fidelity by constraining how low the emission probability can be tuned to reduce multiphoton emissions (Sec. \ref{sec:excessnoise}).  Finally, analysis of the multiplexing factor required to achieve high-fidelity, quasi-deterministic operation with realistic detectors is given in Sec. \ref{sec:multiplexing}.

\section{Heralding entangled pairs from parametric down-conversion}\label{sec:heraldedeps}

\subsection{Swap-heralded dual-TMSV pair source}\label{sec:swapheralded}

We start by reviewing the swap-heralded cascaded SPDC source based on a pair of polarization-entangled SPDC sources heralded by a partial BSM.  A polarization-entangled SPDC source consists of two SPDC processes engineered to produce a TMSV state in a pair of orthogonal modes $\{a_s,b_i\}$ and $\{b_s,a_i\}$, respectively.  For concreteness, we refer to the modes $a,b$ as polarization modes although the following analyses shall apply to any dual-rail basis.  The subscript denotes the spatial mode, referred to as signal and idler modes, and signifies that the modes are otherwise indistinguishable in all other degrees of freedom.  A simplified Hamiltonian describing the interaction between a classical pump field and the down-converted modes is modeled in the form \cite{LAMASLINARES2001}
\begin{equation}\label{eq:dtmsvhamiltonian}
H_{\textup{int}}^{(2)} = e^{i\phi}\kappa (a_s^\dagger b_i^\dagger + b_s^\dagger a_i^\dagger) + \textup{\emph{h.c.}}
\end{equation}
where $a_s^\dagger b_i^\dagger + b_s^\dagger a_i^\dagger$ is the creation operator for entangled photon pairs in the signal and idler channels, $\kappa$ is a real-valued coupling coefficient, and $\phi$ is a phase (without loss of generality, we shall henceforth set $\phi= 0$).  Acting on the vacuum state in the signal and idler modes, the time evolution operator $\hat U(t) = e^{-iH_{\textup{int}}^{(2)}t/\hbar}$ yields a state we shall refer to as a dual-TMSV (D-TMSV) state
\begin{equation}\label{eq:dtmsv}
|\psi\rangle = \sum_{n=0}^\infty \frac{\tanh^n (r)}{\cosh^2(r)} \sum_{k=0}^n |n-k,k\rangle_s|k,n-k\rangle_i
\end{equation}
where $|m,n\rangle$ denotes a Fock state in modes $\{a,b\}$ and the dimensionless interaction parameter $r = \kappa t/\hbar$ is determined by the coupling coefficient $\kappa$ and interaction time $t$.  The probability of producing $n$ total photon pairs from a D-TMSV state is thus given by
\begin{equation}\label{eq:dtmsvpairprob}
p_2(n) = (1-\lambda)^2(n+1)\lambda^{n}
\end{equation}
where $\lambda = \tanh^{2}(r)$ determines the mean photon number per mode via the relation $\mu=\sinh^2(r)=\lambda/(1-\lambda)$.  The subscript on \eqref{eq:dtmsvpairprob} denotes the number of TMSV processes and is used for consistency with later notation.

The D-TMSV state \eqref{eq:dtmsv} should be understood as nothing more than a tensor product of two independent TMSV states $|\psi\rangle = |\textup{TMSV}\rangle|\overline{\textup{TMSV}}\rangle$---where the signal and idler polarizations are swapped on the second factor---as it can be rewritten in the form
\begin{align}\label{eq:tmsv2}
\begin{split}
|\psi\rangle = \sum_{\ell=0}^\infty \frac{\tanh^\ell (r)}{\cosh(r)}|\ell_a\rangle_{s}|\ell_b\rangle_{i} \sum_{k=0}^\infty \frac{\tanh^k (r)}{\cosh(r)}|k_b\rangle_{s}|k_a\rangle_{i},
\end{split}
\end{align}
recovering the former expression via a simple change of summation index $n=\ell+k$.  Nevertheless, we sometimes refer to expression \eqref{eq:dtmsv} to emphasize the grouping by total photon number $n$ and the confinement of emissions onto two polarizations of just two spatial modes.  The latter can be achieved either by combining the output of two type-II SPDC processes on a polarizing beam-splitter (PBS)---e.g., by bidirectionally pumping the source crystal in a Sagnac interferometer configuration \cite{KIM2006}---or by engineering a periodically poled crystalline waveguide with a dual quasi-phase matching structure to simultaneously support both processes \cite{LEVINE2011}.

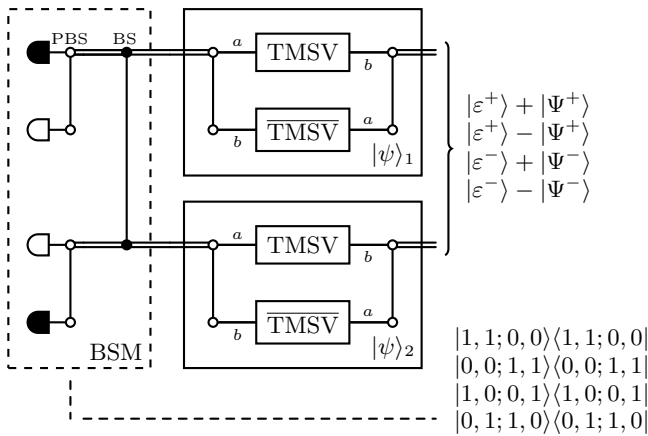
\begin{figure}[h!]
\begin{quantikz}[row sep=0.5cm, column sep=0.5cm]
\inputD[style={fill=black}]{}\gategroup[5,steps=4,style={dashed,inner ysep=6pt,inner xsep=4pt,xshift=0.0em}, label style={below,yshift=-14.0em,xshift=1.5em}]{BSM} & [-0.3cm]
\ctrl[open]{1} \setwiretype{q} & [-0.4cm]
\setwiretype{c} \wire[l][1]["\textup{PBS}"{above,pos=1.5}]{c}\wire[r][1]["\textup{BS}"{above,pos=1.05}]{c}& 
\ctrl{3} 
&
&
\ctrl[open]{1} \gategroup[2,steps=5,style={inner sep=6pt}, label style={below,yshift=-5.8em,xshift=3.7em}]{$|\psi\rangle_1$} & [-0.5cm]
\setwiretype{q} \wire[r][1]["a"{above,pos=0.5}]{q} &
\gate{\textup{TMSV}} &
\wire[l][1]["b"{below,pos=0.5}]{q} & [-0.5cm]
\ctrl[open]{1} &
\setwiretype{c} \rstick[4]{$\begin{array}{c}
|\varepsilon^+\rangle + |\Psi^+\rangle \\ 
|\varepsilon^+\rangle - |\Psi^+\rangle \\ 
|\varepsilon^-\rangle + |\Psi^-\rangle \\ 
|\varepsilon^-\rangle - |\Psi^-\rangle
\end{array}$} \\
\inputD{} &
\control[open]{} &
\setwiretype{n} & 
&
&
\control[open]{} & 
\setwiretype{q} \wire[r][1]["b"{below,pos=0.5}]{q} & 
\gate{\overline{\textup{TMSV}}} &
\wire[l][1]["a"{above,pos=0.5}]{a} &
\control[open]{} \\ 
\\
\inputD{} &
\ctrl[open]{1}&
\setwiretype{c} &
\control{} & 
&
\ctrl[open]{1} \gategroup[2,steps=5,style={inner sep=6pt}, label style={below,yshift=-5.8em,xshift=3.7em}]{$|\psi\rangle_2$} & 
\setwiretype{q} \wire[r][1]["a"{above,pos=0.5}]{q} &
\gate{\textup{TMSV}} &
\wire[l][1]["b"{below,pos=0.5}]{q} &
\ctrl[open]{1} & 
\setwiretype{c} \\
\inputD[style={fill=black}]{} &
\control[open]{} &
\setwiretype{n} & 
&
&
\control[open]{} &
\setwiretype{q} \wire[r][1]["b"{below,pos=0.5}]{q} &
\gate{\overline{\textup{TMSV}}} &
\wire[l][1]["a"{above,pos=0.5}]{a} & 
\control[open]{} \\ 
& \setwiretype{n}\wire[d][1][style={dashed}]{q} & & & & & & & & & \rstick[1]{$\begin{array}{l}
|1,1;0,0\rangle\langle 1,1; 0,0| \\
|0,0;1,1\rangle\langle 0,0; 1,1| \\
|1,0;0,1\rangle\langle 1,0; 0,1| \\
|0,1;1,0\rangle\langle 0,1; 1,0| \end{array}$} \\
& \setwiretype{n} & \setwiretype{q}\wire[r][8][style={dashed}]{q}\setwiretype{n} & & & & & & & & 
\end{quantikz}
\captionsetup{font=footnotesize,labelfont=footnotesize,justification=raggedright}
\caption{Swap-heralded pair source based on a partial BSM between two polarization-entangled SPDC sources, each producing a D-TMSV state.   Solid horizontal lines represent optical modes, with the signal (idler) modes always drawn on the left (right). Filled and open circles joined by a vertical line represent beam-splitter (BS) and polarizing beam-splitter (PBS) unitaries between two modes.  The heralded states $|\varepsilon^\pm\rangle\pm|\Psi^\pm\rangle$ are described in \eqref{eq:bellstates}-\eqref{eq:errorterms}.}
\label{fig:swapheraldedeps}
\end{figure}

For certain applications, rather than using the single-pair emissions directly as unheralded entangled pairs, it may be desirable to filter the vacuum and multi-pair terms using a heralding signal. The swap-heralded source \cite{DHARA2022} achieves this by combining two D-TMSV sources in an entanglement swapping protocol illustrated in Fig. \ref{fig:swapheraldedeps}.
In this scheme, two D-TMSV states are mixed by combining the signal modes of the joint state $|\psi\rangle_1|\psi\rangle_2$ on a beam-splitter followed by detection of the output modes
\small
\begin{align}\label{eq:bs1}
&\tilde a_{1,s} = \sqrt{1/2}(a_{1,s} + a_{2,s}) 
&\tilde a_{2,s} = \sqrt{1/2}(a_{1,s} - a_{2,s}) \\
&\tilde b_{1,s} = \sqrt{1/2}(b_{1,s} + b_{2,s}) 
&\tilde b_{2,s} = \sqrt{1/2}(b_{1,s} - b_{2,s}).
\label{eq:bs2}
\end{align}
\normalsize
A probabilistic measurement of exactly two photons of opposite polarizations $\tilde a_{j,s}\tilde b_{k,s}$ in the output modes heralds a successful entanglement swap, with the goal of producing a Bell pair in the idler modes.   

More precisely, in the absence of losses and assuming perfect detection of the heralding modes, the unnormalized component of the state $|\psi\rangle_1|\psi\rangle_2$ which can yield a successful heralding event takes the form
\small
\begin{align}\label{eq:swapheraldedstate1}
&\frac{1}{2}\tilde a_{1,s}^\dagger \tilde b_{1,s}^\dagger\left(  a_{1,i}^\dagger b_{1,i}^\dagger + a_{2,i}^\dagger b_{2,i}^\dagger + [a_{1,i}^\dagger b_{2,i}^\dagger + b_{1,i}^\dagger a_{2,i}^\dagger] \right)| 0 \rangle  \\
&+\frac{1}{2}\tilde a_{2,s}^\dagger \tilde b_{2,s}^\dagger\left(a_{1,i}^\dagger b_{1,i}^\dagger + a_{2,i}^\dagger b_{2,i}^\dagger - [a_{1,i}^\dagger b_{2,i}^\dagger + b_{1,i}^\dagger a_{2,i}^\dagger] \right)| 0 \rangle \\
&+\frac{1}{2}\tilde a_{1,s}^\dagger \tilde b_{2,s}^\dagger\left(  a_{1,i}^\dagger b_{1,i}^\dagger - a_{2,i}^\dagger b_{2,i}^\dagger + [a_{1,i}^\dagger b_{2,i}^\dagger - b_{1,i}^\dagger a_{2,i}^\dagger] \right)| 0 \rangle \\
&+\frac{1}{2}\tilde a_{2,s}^\dagger \tilde b_{1,s}^\dagger\left(  a_{1,i}^\dagger b_{1,i}^\dagger - a_{2,i}^\dagger b_{2,i}^\dagger - [a_{1,i}^\dagger b_{2,i}^\dagger - b_{1,i}^\dagger a_{2,i}^\dagger] \right)| 0 \rangle
\label{eq:swapheraldedstate4}
\end{align}
\normalsize
where we have left off the factor $\tanh^2(r)/\cosh^4(r)$.  The terms in square brackets
\begin{equation}\label{eq:bellstates}
|\Psi^\pm\rangle =  \sqrt{1/2}[a_{1,i}^\dagger b_{2,i}^\dagger \pm b_{1,i}^\dagger a_{2,i}^\dagger]|0\rangle
\end{equation}
correspond to a desired Bell state of the swap-heralded source, and the remaining terms
\begin{equation}\label{eq:errorterms}
|\varepsilon^\pm\rangle =  \sqrt{1/2}\big( a_{1,i}^\dagger b_{1,i}^\dagger \pm a_{2,i}^\dagger b_{2,i}^\dagger \big)|0\rangle
\end{equation}
represent errors arising when one of the two D-TMSV sources emits a double-pair (processes for filtering the error terms are discussed in Sec. \ref{sec:noonswap}).  

The efficiency of heralding events from the swap-heralded source is driven by the probability of two-pair emissions in the joint state $|\psi\rangle_1|\psi\rangle_2$.  The probability of producing $n$ total pairs is given by
\begin{equation}\label{eq:dtmsv2pairprob}
p_4(n) = \frac{1}{6}(1-\lambda)^4(n+1)(n+2)(n+3)\lambda^n.
\end{equation}
The two-pair emissions are maximized with $\lambda = 1/3$ corresponding to a mean photon number per mode of $\mu=1/2$.  
A straightforward calculation reveals that 2/5 of the 4-photon emissions yield oppositely-polarized signal photons resulting in one of the four successful heralding events described by \eqref{eq:swapheraldedstate1}-\eqref{eq:swapheraldedstate4}, yielding a maximum possible heralding probability
\begin{equation}\label{eq:doubleprobtmsv2}
P_{2}^{(2)} = (2/5)p_4(2) \leq  (2/3)^6\simeq 0.088.
\end{equation}

Of course, this theoretical maximum is of limited practical interest, since realistic detectors constrain the mean photon number per mode significantly below $\mu=1/2$ in order to reduce higher-order emissions which can produce erroneous heralding events in the presence of losses.  A proper analysis of the efficiency and fidelity of the source requires consideration of the detector POVM (positive operator-valued measure) characterizing the measurement performed by the heralding detectors (Sec. \ref{sec:realistic}).

\subsection{Double-heralded TMSV pair source}\label{sec:doubleheralded}

\begin{figure}[h!]
\begin{quantikz}[row sep=0.25cm, column sep=0.3cm]
\lstick[2]{$|1,1\rangle\langle 1,1|$}&
\inputD[style={fill=black}]{} & 
\wire[r][1]["a"{above,pos=0.5}, draw=none]{q} &
\gate{\textup{TMSV}} &
\wire[l][1]["b"{below,pos=0.5}, draw=none]{q} & 
&
\ctrl[open]{1} &
 \setwiretype{c}\rstick[1]{$a_1^\dagger b_1^\dagger |0\rangle\qquad\quad$} \\
&
\inputD[style={fill=black}]{} & [-0.5cm]
\wire[r][1]["b"{below,pos=0.5}, draw=none]{q} & 
\gate{\overline{\textup{TMSV}}} &
\wire[l][1]["a"{above,pos=0.5}, draw=none]{a} &
&
\control[open]{} &
\setwiretype{n}
\end{quantikz}
$$(a)$$
\vspace{-25pt}
\begin{center}
\tikz{\draw[dashed] (0,0) -- (8,0);}
\end{center}
\begin{quantikz}[row sep=0.25cm, column sep=0.3cm]
\lstick[2]{$|1,1\rangle\langle 1,1|$}&
\inputD[style={fill=black}]{} &
\wire[r][1]["a"{above,pos=0.5}, draw=none]{q} &
\gate{\textup{TMSV}} &
\wire[l][1]["b"{below,pos=0.5}, draw=none]{q} & 
&
\ctrl[]{1} &
\setwiretype{c}\rstick[2]{$
|\varepsilon^-\rangle + |\Psi^-\rangle$} \\
&
\inputD[style={fill=black}]{} & [-0.5cm]
\wire[r][1]["b"{below,pos=0.5}, draw=none]{q} & 
\gate{\overline{\textup{TMSV}}} &
\wire[l][1]["a"{above,pos=0.5}, draw=none]{a} &
&
\control[]{} & \setwiretype{c}
\end{quantikz}
$$(b)$$
\vspace{-25pt}
\begin{center}
\tikz{\draw[dashed] (0,0) -- (8,0);}
\end{center}
\begin{quantikz}[row sep=0.25cm, column sep=0.3cm]
\lstick[2]{$|1,1\rangle\langle 1,1|$}&
\inputD[style={fill=black}]{} &
\wire[r][1]["a"{above,pos=0.5}, draw=none]{q} &
\gate{\textup{TMSV}} &
\wire[l][1]["b"{below,pos=0.5}, draw=none]{q} &
&
\push{\scriptstyle\otimes}
\wire[d][1]["\times\textup{-PBS}"{above,pos=0.0,yshift=0.15cm}]{q} &
\setwiretype{c}\rstick[2]{$|\varepsilon_{cd}^-\rangle + |\Psi_{cd}^-\rangle$} \\
&
\inputD[style={fill=black}]{} & 
\wire[r][1]["a"{above,pos=0.5}, draw=none]{q} & 
\gate{\textup{TMSV}} &
\wire[l][1]["b"{below,pos=0.5}, draw=none]{a} &
&
\push{\scriptstyle\otimes} & \setwiretype{c}
\end{quantikz}
$$(c)$$
\vspace{-15pt}
\captionsetup{font=footnotesize,labelfont=footnotesize,justification=raggedright}
\caption{Double-heralded pair source configurations based on 4-photon emissions from two TMSV sources, heralded by two single-photon detections on the signal modes.  The $\times$-PBS denotes a diagonally oriented polarizing beam-splitter which produces a partial Bell state in the basis $c=\sqrt{1/2}(a+b)$, $d=\sqrt{1/2}(a-b)$.}
\label{fig:doubleheraldedeps}
\end{figure}
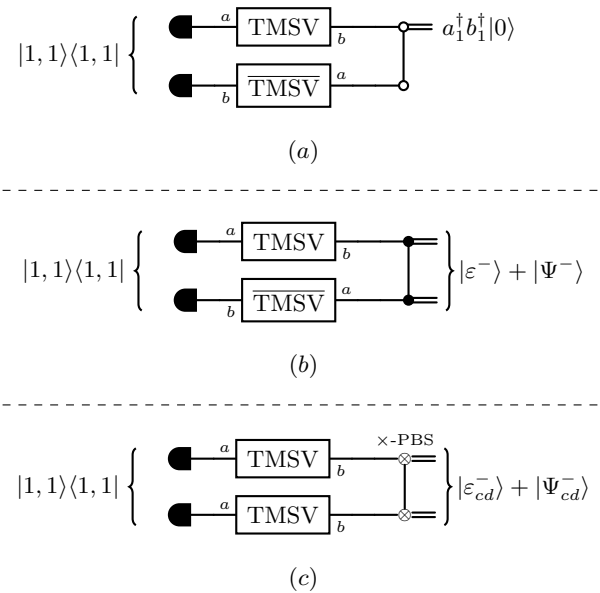

In this section, we present a simplification of the swap-heralded dual-TMSV source above which eliminates the need for an entanglement swap between two polarization-entangled sources and simplifies the optics needed on the heralding path.  Specifically, consideration of the complete heralded state in the idler modes of \eqref{eq:swapheraldedstate1}-\eqref{eq:swapheraldedstate4}, including non-Bell-pair terms, shows that each state is unitarily equivalent to a simple anti-correlated pair
\begin{equation}\label{eq:anticorrelatedpair}
a^{\dagger}b^\dagger|0\rangle.
\end{equation}
This equivalence is realized by placing the anti-correlated pair through a 50:50 beam-splitter \eqref{eq:bs1}-\eqref{eq:bs2}, e.g.
\small
\begin{equation}\label{eq:dtmsvheraldedstate}
a_i^\dagger b_i^\dagger |0\rangle = \frac{1}{2}\Big( \tilde a_{1,i}^\dagger \tilde b_{1,i}^\dagger + \tilde a_{2,i}^\dagger \tilde b_{2,i}^\dagger + [\tilde a_{1,i}^\dagger \tilde b_{2,i}^\dagger + \tilde b_{1,i}^\dagger \tilde a_{2,i}^\dagger] \Big)|0\rangle
\end{equation}
\normalsize
with the other three heralded states obtained from each possible choice of input ports for the two photons.  The Bell pair term in brackets arises from the case where the photons exit different output ports.  The swap-heralded source can hence be understood simply as a source of anti-correlated photon pairs.  In particular, the full apparatus of Fig. \ref{fig:swapheraldedeps} is not required, as one can instead produce anti-correlated pairs by heralding single-photons from independent TMSV sources.  Two heralded single photons from a double-heralding event can then be combined via linear optics into the desired form, as shown in Fig. \ref{fig:doubleheraldedeps}.

Observe that distinct TMSV sources need not even produce photons of orthogonal polarizations, as the form of the two-photon state in \eqref{eq:anticorrelatedpair} is really quite general, and simply requires two photons in orthogonal modes.  For example, Fig. \ref{fig:doubleheraldedeps}($c$) represents a TMSV configuration with two identical TMSV states heralding photons in polarization mode $b$ which are then combined on a polarizing beam-splitter (PBS) oriented diagonally to produce a partial Bell pair. This point of view, \emph{i.e.}, understanding the swap-heralded source simply as a source of anti-correlated pairs, greatly broadens the possibilities for implementations of ZALM and cascaded SPDC to suit a variety of engineering constraints.

Despite the formal equivalence of the heralded states produced by the swap-heralded and double-heralded sources, there are a number of subtle but important differences.  First, the double-heralded source reduces the amount of mode-sorting optics needed on the heralding path, improving the heralding efficiency which is critical to the fidelity of the source.  However, this difference is not fundamental, as we shall show in the next section that the swap-heralded source can be reconfigured into a double-heralded configuration to minimize heralding losses.  The fundamental difference between the two schemes is that the swap-heralded source requires detection of exactly two photons in four modes as opposed to two photons in two modes.  As a result, the schemes exhibit subtle differences in pair generation statistics.  In the double-heralded TMSV configuration, the probability with which higher order pairs are produced is governed by the TMSV pair generation probability
\begin{equation}
p_1(n) = (1-\lambda)\lambda^n
\end{equation}
in place of \eqref{eq:dtmsvpairprob}.  In particular, the probability with which single-pairs are produced by independent TMSV processes is maximized with $\lambda=1/2$, corresponding to a mean photon number per mode $\mu=1$, and the maximum probability of a double-heralding event consisting of one pair from each of two TMSV sources is
\begin{equation}\label{eq:doubleprobtmsv1}
P_{1}^{(2)} = p_1(1)^2\leq\left(1/2\right)^4= 0.0625.
\end{equation} 

Note that the maximum heralding probability $P_{2}^{(2)}= 0.088$ from the swap-heralded D-TMSV sources is larger than that from the double-heralded TMSV sources by a factor of about $1.4$.  This gain is effectively a multiplexing gain, since it comes at the cost of requiring twice as many detectors and sources together with active feed-forward operations to handle the 4 valid click patterns in the partial BSM.  In this regard, it is notable that the use of D-TMSV sources in place of TMSV sources only provides a gain of roughly $ \sqrt{2}$ with its mode multiplexing in the quantum domain, rather than the full factor of 2 improvement in pair heralding rate one might expect when employing twice as many TMSV sources and detectors.  This reduced gain from the additional sources is accounted for by noting that the optimal mean photon \emph{per mode} $\mu=1/2$ for generating a single pair from a D-TMSV source is half that of a TMSV source, since the former employs twice as many TMSV sources (and modes) to produce a pair.

Of course, we must again emphasize that with realistic detectors, the mean photon number per mode $\mu$ must be reduced to limit higher order pairs at the cost of a potentially significant reduction in the efficiency with which double-pairs are generated. A complete analysis of this trade-off with realisitic detectors is given in Sec. \ref{sec:realistic}.

\subsection{Double-heralded multimode TMSV sources}\label{sec:doubleheraldedmtmsv}

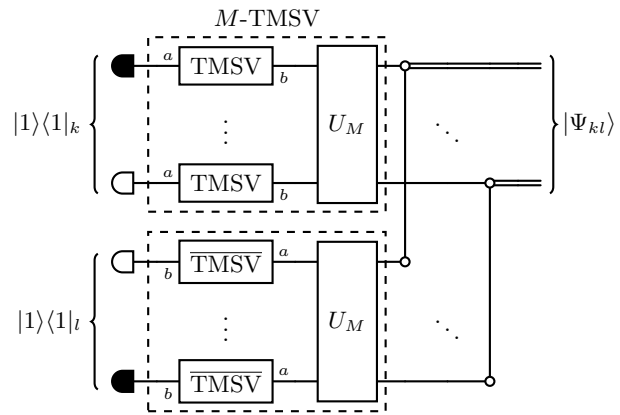
\begin{figure}[h!]
\begin{quantikz}[row sep=0.25cm, column sep=0.3cm]
\lstick[3]{$
|1\rangle\langle 1|_k$}&[-0.25cm]
\inputD[style={fill=black}]{} &
\gategroup[3,steps=4,style={dashed,inner ysep=0pt,inner xsep=0pt}, label style={}]{$M$-TMSV}
\wire[r][1]["a"{above,pos=0.5}, draw=none]{q} &
\gate{\textup{TMSV}} &
\wire[l][1]["b"{below,pos=0.5}, draw=none]{q} & 
\gate[3]{U_M} &
\ctrl[open]{4} & \setwiretype{c} & 
&
&
 \rstick[3]{$|\Psi_{kl}\rangle$} 
\\
\setwiretype{n} & & & \vdots & & & & \ddots & &
\\
&
\inputD{} &
\wire[r][1]["a"{above,pos=0.5}, draw=none]{q} & 
\gate{{\textup{TMSV}}} &
\wire[l][1]["b"{below,pos=0.5}, draw=none]{a} &
&
&
&
\ctrl[open]{4} &
\setwiretype{c} 
&
\\  
\\
\lstick[3]{$|1\rangle\langle 1|_l$}&
\inputD{} &
\gategroup[3,steps=4,style={dashed,inner ysep=0pt,inner xsep=0pt}, label style={}]{}
\wire[r][1]["b"{below,pos=0.5}, draw=none]{q} &
\gate{\overline{\textup{TMSV}}} &
\wire[l][1]["a"{above,pos=0.5}, draw=none]{q} & 
\gate[3]{U_M} &
\control[open]{} &
\setwiretype{n} &
&
&
\\
\setwiretype{n} & & & \vdots & & & & \ddots &
\\
&
\inputD[style={fill=black}]{} & [-0.5cm]
\wire[r][1]["b"{below,pos=0.5}, draw=none]{q} & 
\gate{\overline{\textup{TMSV}}} &
\wire[l][1]["a"{above,pos=0.5}, draw=none]{a} &
&
&
&
\control[open]{}
\end{quantikz}
\captionsetup{font=footnotesize,labelfont=footnotesize,justification=raggedright}
\caption{Double-heralded $M$-TMSV configurations based on a 2-photon emission from each of two $M$-TMSV arrays, heralded by a single photon detection on the signal modes from each array.  The operator $U_M$ represents an $M$-mode QFT or Hadamard unitary. For $M=2$, $U_2$ is just a 50:50 beam-splitter and this configuration is formally equivalent to the swap-heralded configuration of Fig. \ref{fig:swapheraldedeps} (Appendix \ref{app:symmetry}).}
\label{fig:doubleheraldedmultimode}
\end{figure}

In this section we introduce another generalization of the swap-heralded D-TMSV scheme, motivated by the observation that using D-TMSV sources reduces the optimal mean photon number per mode and increases the pair heralding probability.  First, we can rearrange the swap-heralded source into a double-heralded configuration to simplify the heralding path, as shown in Fig. \ref{fig:doubleheraldedmultimode}.  This follows from a symmetry of the $M$-fold TMSV Hamiltonian $H_{\textup{int}}^{(M)}$, whereby any unitary applied to the signal modes can instead be applied to the idler modes to produce an equivalent state (Appendix \ref{app:symmetry}).

Each dashed box represents a multimode $M$-TMSV array consisting of $M$ identical TMSV sources, with heralding detectors directly on the signal modes and an $M$-port interferometer applied to the idler modes.  For $M=1$ this is just the configuration of Fig. \ref{fig:doubleheraldedeps}(a).  For $M=2$, the double-heralded 2-TMSV configuration is formally equivalent to the swap-heralded D-TMSV configuration of Fig. \ref{fig:swapheraldedeps} via the identity in Appendix \ref{app:symmetry}, but reduces the optics required on the heralding path.  Each multimode $M$-TMSV array \emph{jointly} heralds the production of a single photon, i.e., we define the heralding protocol such that a successful herald from an $M$-TMSV array consists of a detection of exactly one photon, with all other detectors registering vacuum.  This projects the idler modes of each array onto a state consisting of a single photon superposed on $M$ spatial modes at the output of an $M$-port interferometer $U_M$.

A double-heralding event from two such arrays---producing orthogonally polarized idler photons---can be reduced to an anti-correlated pair state superposed equally on $M$ spatial modes
\begin{equation}
|\Psi_{kl}\rangle=\frac{1}{M}\sum_{i,j=1}^M e^{i\phi_{ij}} a_i^\dagger b_j^\dagger|0\rangle
\end{equation}
where the phase of each term is determined by the two detectors (labeled $k$ and $l$) that registered a click.  For example, using a quantum Fourier transform (QFT) unitary on the idler modes, the phases are determined by the QFT matrix, or if $M$ is a power of 2, the phase factors can be taken to be $\pm 1$ using an $M$-port Hadamard unitary.  Each spatial mode can be distributed to a different receiver in a star topology with the heralded multimode source at the central node, and entanglement distribution succeeds when two different receivers each herald loading of a single photon into a quantum memory, where a phase correction can be applied to recover a specified Bell state.  For $M\geq 2$, given two jointly-heralded photons, entanglement distribution succeeds with maximum probability
\begin{equation}
\mathcal P(\textup{Bell}|\Psi) = \frac{M-1}{M}
\end{equation}
reducing to the 1/2 success probability of entanglement distribution from the swap-heralded source for $M=2$.  Thus, the efficiency of the $M$-TMSV source as an entanglement distribution node improves with the number of receiver nodes.

Observe that for $M=1$, since each pair is produced in a single spatial mode, the pair can be deterministically split equally into $M$ spatial modes via a 1 to $M$ splitter, yielding the same maximum success probability for entanglement distribution.  However, each $M$-TMSV array yields different pair generation statistics depending on the number of modes $M$, motivating a general analysis of the $M$-TMSV source.

The probability of producing a single pair from an array of $M$ independent TMSV sources is
\begin{equation}\label{eq:singleheraldprobability}
p_M\equiv p_M(1) = M p_1(1)p_1(0)^{M-1} = M(1-\lambda)^{M}\lambda,
\end{equation}
where we abuse our notation by identifying $p_M$ as $p_M(1)$.  This pair probability is maximized when $\lambda = 1/(M+1)$, with optimal mean photon number per mode
\begin{equation}\label{eq:optimalmumtmsv}
\hat\mu = 1/M.
\end{equation} 
The maximal probability of a double-herald event generating a single photon from each of two $M$-TMSV arrays is then
\begin{equation}\label{eq:truepairgenerationideal}
P_{M}^{(2)} = p_M^2 \leq \left(\frac{M}{M+1}\right)^{2M+2}
\end{equation}
which reduces to \eqref{eq:doubleprobtmsv1} and \eqref{eq:doubleprobtmsv2} for $M=1$ and $M=2$, respectively, and limits to $1/e^2\simeq 0.135$ as $M\to \infty$.  In effect, the multimode TMSV array behaves as a photon source with improved single-pair generation statistics.  However, $M$-TMSV arrays yield quickly diminishing gains in terms of maximum pair generation rate relative to the number of sources and detectors.

The fundamental difference between the double-heralded TMSV source and swap-heralded D-TMSV source, understanding that the swap-heralding protocol can be replaced with a double-heralding protocol, can hence be understood as the difference between generating photons from multimode $M$-TMSV photon sources with $M=1$ and $M=2$, respectively.  Given a fixed set of resources (sources and detectors), the comparison of the overall performance of the two schemes can thus be understood as a comparison between resources and implementation constraints of multiplexed $M$-TMSV sources.

\subsection{Enhanced pair generation probability through multiplexing with active mode-switching}\label{sec:doubleheralding}

\begin{figure}[h!]
\begin{quantikz}[row sep=0.25cm, column sep=0.25cm]
&[-0.4cm]
\setwiretype{n}\wire[d][4][style={dotted}]{q}
\wire[r][13][style={dotted}]{q}&
 & & & & & &
\ctrl[style={dotted}]{1}& 
\ctrl[style={dotted}]{2}&
\ctrl[style={dotted}]{2}&
&
\ctrl[style={dotted}]{4} &
\ctrl[style={dotted}]{4} &
\rstick{$(k,l)$}
\\
&
\setwiretype{q}\wire[r][1][style={dotted}]{q}\setwiretype{n}&
\inputD[style={fill=black}]{}&
\setwiretype{q} \qwbundle{M} &
\wire[r][1]["a"{above,pos=0.5}, draw=none]{q}&
\gate{\textup{$M$-TMSV}} &
\wire[l][1]["b"{below,pos=0.5}, draw=none]{q}\qwbundle[style={xshift=0.25cm}]{M} & 
&
\octrl{4} & 
\setwiretype{n} & 
&
&
\\
&
\setwiretype{q}\wire[r][1][style={dotted}]{q}\setwiretype{n}&
\inputD[style={fill=black}]{} &
\setwiretype{q} \qwbundle{M} &
\wire[r][1]["a"{above,pos=0.5}, draw=none]{q} & 
\gate{M\textup{-TMSV}} &
\wire[l][1]["b"{below,pos=0.5}, draw=none]{a}\qwbundle[style={xshift=0.25cm}]{M} &
&
&
\gate{X} &
\ctrl[open]{3} &
\setwiretype{n} &
\\  
\setwiretype{n}& & & & & \vdots & & & & & & \ddots &
\\
&
\setwiretype{q}\wire[r][1][style={dotted}]{q}\setwiretype{n}&
\inputD[style={fill=black}]{} &
\setwiretype{q} \qwbundle{M} &
\wire[r][1]["a"{above,pos=0.5}, draw=none]{q} &
\gate{{\textup{$M$-TMSV}}} &
\wire[l][1]["b"{below,pos=0.5}, draw=none]{q}\qwbundle[style={xshift=0.25cm}]{M} & 
&
&
&
&
&
\gate{X}&
\octrl{1}&
\setwiretype{n}
\\
&\setwiretype{n} & & & & & & & \ocontrol{} & \setwiretype{q}\qwbundle{M} &
\ocontrol{} &
\setwiretype{c} & &
\ocontrol{} &
\rstick[1]{$|\Psi_{kl}\rangle$}
\end{quantikz}
\captionsetup{font=footnotesize,labelfont=footnotesize,justification=raggedright}
\caption{Double-heralded $M$-TMSV in a complete multiplexing configuration with active mode-conversion into a prescribed set of $M$ modes for injection into a network. If at least two $M$-TMSV sources produce a pair, the corresponding idler modes are switched into the output channel with one photon of each polarization superposed on $M$ modes.  Dotted lines represent classical control signals. The controlled-$X$ gate represents a half-wave plate swapping polarizations, and the controlled-PBS selectively switches the modes which contain a heralded photon into the output channel.  The indices $1\leq k,l\leq M$ of the detectors which registered a click are distributed with the anti-correlated pair to apply the corresponding phase shifts.}
\label{fig:multiplexedmtmsv}
\end{figure}
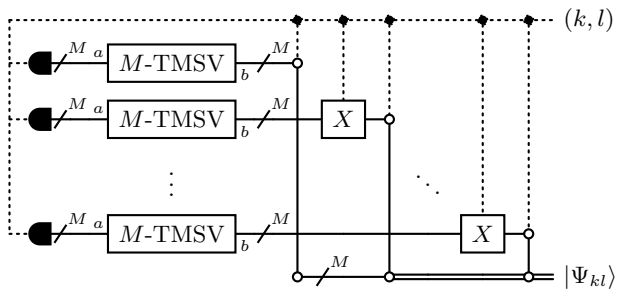

We now turn to a consideration of the enhancement of the pair generation probability through multiplexing, relying on active mode-conversion to produce pairs superposed on a predefined set of $M$ dual-rail channels.  Understanding each $M$-TMSV source as a heralded source of single photons, they can be multiplexed simply by determining which two $M$-TMSV sources heralded a single-photon, and employing active mode-switching to load the photon pair onto the output channels, as in Fig. \ref{fig:multiplexedmtmsv}.

Given the probability $p_M$ that an $M$-TMSV array heralds a single photon, the probability that at least 2 of $N$ such arrays produces a photon is given by
\begin{equation}\label{eq:multiplexedpgp}
P_M^{(2)}(N) = 1 - (1-p_M)^N - Np_M(1-p_M)^{N-1}.
\end{equation}
Assuming that the multiplexed source emits at most one heralded pair per pump pulse, this represents the pair generation probability of $N$ multiplexed $M$-TMSV sources.  This expression yields an equation for the number of multiplexed sources $N$ needed to achieve quasi-deterministic operation targeting pair generation with efficiency $\eta_h=P_M^{(2)}(N)$.  The solution can be expressed as
\begin{equation}\label{eq:maximalmultiplexingrequirement}
N(\eta_h) = \frac{W\big((1-\eta_h)(1-q)\log(q)q^{1-q}\big)}{\log(q)}-\frac{q}{1-q},
\end{equation}
where $q = 1-p_M$ and $W$ is the Lambert $W$ function.

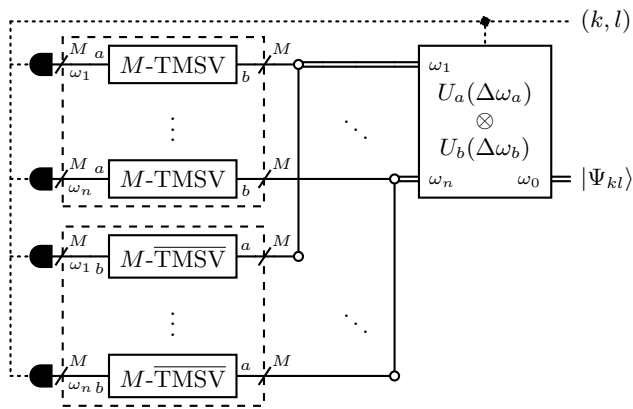
\begin{figure}[h!]
\begin{quantikz}[row sep=0.25cm, column sep=0.25cm]
&[-0.4cm]
\setwiretype{n}\wire[d][7][style={dotted}]{q}
\wire[r][12][style={dotted}]{q}&
 & & & & & &
& 
&
&
&
\ctrl[style={dotted}]{1}&
\rstick{$(k,l)$}
\\
&
\setwiretype{q}\wire[r][1][style={dotted}]{q}\setwiretype{n}&
\inputD[style={fill=black}]{}&
\gategroup[3,steps=4,style={dashed,inner ysep=0pt,inner xsep=0pt}, label style={}]{}
\setwiretype{q} \qwbundle{M} &
\wire[r][1]["a"{above,pos=0.5}, draw=none]{q}\wire[r][1]["\omega_1"{below,pos=-0.5}, draw=none]{q}&
\gate{\textup{$M$-TMSV}} &
\wire[l][1]["b"{below,pos=0.5}, draw=none]{q}\qwbundle[style={xshift=0.25cm}]{M} & 
&
\octrl{4}& 
\setwiretype{c} & 
&
&
\gate[3,disable auto height]{\begin{array}{c} U_a(\Delta \omega_a) \\ \otimes \\ U_b(\Delta \omega_b) \end{array}}\gateinput{$\omega_1$}
\\
\setwiretype{n}& & & & & \vdots & & & & & \ddots &
\\
&
\setwiretype{q}\wire[r][1][style={dotted}]{q}\setwiretype{n}&
\inputD[style={fill=black}]{} &
\setwiretype{q} \qwbundle{M} &
\wire[r][1]["a"{above,pos=0.5}, draw=none]{q}\wire[r][1]["\omega_n"{below,pos=-0.5}, draw=none]{q} &
\gate{M\textup{-TMSV}} &
\wire[l][1]["b"{below,pos=0.5}, draw=none]{a}\qwbundle[style={xshift=0.25cm}]{M} &
&
&
&
& 
\octrl{4} &
\setwiretype{c}\gateinput{$\omega_n$}\gateoutput{$\omega_0$}&
\rstick{$|\Psi_{kl}\rangle$}
\\ 
\\
&
\setwiretype{q}\wire[r][1][style={dotted}]{q}\setwiretype{n}&
\inputD[style={fill=black}]{}&
\gategroup[3,steps=4,style={dashed,inner ysep=0pt,inner xsep=0pt}, label style={}]{}
\setwiretype{q} \qwbundle{M} &
\wire[r][1]["b"{below,pos=0.5}, draw=none]{q}\wire[r][1]["\omega_1"{below,pos=-0.5}, draw=none]{q}&
\gate{M\textup{-}\overline{\textup{TMSV}}} &
\wire[l][1]["a"{above,pos=0.5}, draw=none]{q}\qwbundle[style={xshift=0.25cm}]{M} & 
&
\ocontrol{} &
\setwiretype{n} & 
&
&
\\
\setwiretype{n}& & & & & \vdots & & & & & \ddots &
\\
&
\setwiretype{q}\wire[r][1][style={dotted}]{q}\setwiretype{n}&
\inputD[style={fill=black}]{} &
\setwiretype{q} \qwbundle{M} &
\wire[r][1]["b"{below,pos=0.5}, draw=none]{q}\wire[r][1]["\omega_n"{below,pos=-0.5}, draw=none]{q} &
\gate{M\textup{-}\overline{\textup{TMSV}}} &
\wire[l][1]["a"{above,pos=0.5}, draw=none]{q}\qwbundle[style={xshift=0.25cm}]{M} & 
&
&
&
&
\ocontrol{}&
\setwiretype{n}
\end{quantikz}
\captionsetup{font=footnotesize,labelfont=footnotesize,justification=raggedright}
\caption{Double-heralded $M$-TMSV in a bipartite multiplexing configuration for frequency-multiplexed ZALM with $n=N/2$ spectral modes. If at least one $M$-TMSV source from each set produces a pair, the corresponding idler modes are switched into the output frequency $\omega_0$ via polarization-dependent frequency-conversion $U_x(\Delta\omega_x)$ which applies a frequency shift $\Delta\omega_x = \omega_0 - \omega_x$ to all $MN/2$ modes with polarization $x$.}
\label{fig:bipartitemultiplexedmtmsv}
\end{figure}

The multiplexing scheme above represents a maximally-connected multiplexing configuration since any two sources can be combined to produce a pair.  This configuration is likely the best configuration for a fully spatially-multiplexed implementation.  Shapiro \emph{et. al.} \cite{SHAPIRO2025,EMBLETON2025} have introduced an alternative bipartite multiplexing scheme for swap-heralded D-TMSV sources which is well-suited for implementation of a ZALM architecture (referred to as cross-island heralding in the context of the frequency-multiplexed implementation, where each D-TMSV pair emits into an ``island" in the joint-spectral intensity of a spectrally-engineered source).  This multiplexing scheme can also be implemented with double-heralded $M$-TMSV arrays as follows (Fig. \ref{fig:bipartitemultiplexedmtmsv}).

First, the $M$-TMSV sources are partitioned into two sets grouped by polarization, one set heralding photons in polarization $a$, and the other set heralding photons with polarization $b$.  Each set consists of $N/2$ frequency modes $\omega_1,...,\omega_{N/2}$.  One $M$-TMSV source from each set emits into each frequency mode.  The two sources which emit into a given frequency mode are paired in a static arrangement that reduces the $2M$ spatial modes to just $M$ total spatial modes (e.g., using polarizing beam-splitters) for distribution to $M$ receivers (for $M=1$, the one resulting spatial mode can be split to two receivers with a 50:50 beam-splitter).  If one photon from each set is heralded, the information describing the frequency-polarization mode-pair of each of the two heralding signals is distributed to the receivers (together with the detection port of each $M$-TMSV array needed to recover the phase), where all of the active mode-switching is done using polarization-dependent frequency-conversion.

The double-heralding succeeds if at least one photon is produced from each set (polarization mode), which occurs with probability
\begin{equation}\label{eq:bipartitepgp}
P_M^{(11)}(N) = \left[1 - (1- p_M)^{N/2}\right]^2
\end{equation}
where the quantity in brackets is the probability that at least one of the $M$-TMSV arrays in a given set heralds a single-photon emission. This relation is easily inverted to obtain the multiplexing factor required to yield quasi-deterministic operation with efficiency $\eta_h$ as 
\begin{equation}
N(\eta_h)=\frac{2\log(1 - \sqrt{\eta_h})}{\log(1 - p_M)}.
\end{equation}
Obviously, the multiplexed pair generation probability \eqref{eq:bipartitepgp} is lower than \eqref{eq:multiplexedpgp} since the possible source combinations used to produce a pair are reduced from a complete graph on $N$ nodes to a complete bipartite graph on $N$ nodes; however, it does offer benefits for the implementation of ZALM.  In particular, by partitioning the sources into two sets by polarization, the receivers can implement polarization-dependent frequency-conversion minimizing switching and mode-sorting losses.  A fully multiplexed source with $N$ frequency modes would require frequency-dependent frequency-conversion, likely incurring additional losses and complexity from the wavelength demultiplexer and resource overhead of independent frequency-converters on each wavelength channel.  On the other hand, the reduction in efficiency from the polarization-partitioned arrangement is bounded by
\begin{equation}
\frac{1}{2}\leq \frac{P_M^{(11)}(N)}{P_M^{(2)}(N)} \leq 1
\end{equation}
with the ratio approaching unity once both sources have become deterministic at $N\sim 8p_M^{-1}$.  

\begin{figure}[h!]
\centering\includegraphics[width=8cm]{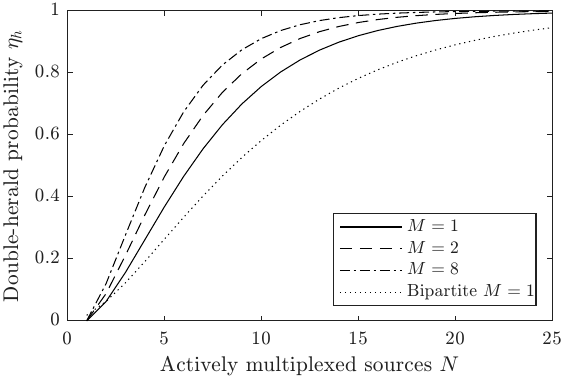}
\captionsetup{font=footnotesize,labelfont=footnotesize,justification=raggedright}
\caption{Double-heralding probability $\eta_h$ for $N$ actively multiplexed $M$-TMSV arrays with lossless PNR detection.  The upper three curves assume a maximally-connected multiplexing configuration.}
\label{fig:multiplexedefficiency}
\end{figure}

Fig. \ref{fig:multiplexedefficiency} shows the probability of a double-heralding event as a function of the number $N$ of actively multiplexed $M$-TMSV sources.  With $M=2$ we find a reduction in the required number of sources needed to achieve a fixed probability $\eta_h$ of $\sim 0.8$, with a similar reduction obtained increasing from $M=2$ to $M=8$.  This reduction is found to be roughly uniform for $\eta_h>0.2$.  Since the total number of sources and detectors is given by the product $MN$, this modest reduction in the required number of \emph{actively} multiplexed sources comes at the cost of twice the number of total sources and detectors for $M=2$, and eight times as many sources and detectors for $M=8$, showing that the double-heralded TMSV with $M=1$ yields the most resource-efficient implementation.  Nevertheless, using multimode $M$-TMSV sources does provide a reduction in the number of multiplexed sources that require fast active mode-switching, since the multimode $M$-TMSV source requires only phase shifts that can be applied after loading the pair in quantum memory.

\subsection{Heralded Bell pairs}\label{sec:noonswap}
An anti-correlated pair of otherwise-indistinguishable photons is unitarily equivalent to a 2-photon N00N state
\begin{equation}\label{eq:noonstate}
a_i^\dagger b_i^\dagger |0\rangle =\frac{1}{2}(c_i^\dagger c_i^\dagger - d_i^\dagger d_i^\dagger)|0\rangle
\end{equation}
in the diagonal basis 
\begin{align}\label{eq:diagonalbasis}
c_i &= \sqrt{1/2}(a_i + b_i) \\
d_i &= \sqrt{1/2}(a_i - b_i)
\end{align}
and has a number of potential uses in quantum information and sensing applications.  Indeed, configurations similar to Fig. \ref{fig:doubleheraldedeps} have been previously demonstrated primarily as sources of 2-photon N00N states \cite{EISENBERG2005,SMITH2008}.  However, our primary motivation for the heralded sources above is as a source of Bell pairs.  One caveat is that the raw fidelity of the heralding schemes above as a source of Bell pairs is limited to 50\% due to the presence of the erroneous two-photon terms 
\begin{align}\label{eq:errortermab}
|\varepsilon^+\rangle &= \sqrt{1/2}\big(a_{1,i}^\dagger b_{1,i}^\dagger + a_{2,i}^\dagger b_{2,i}^\dagger \big)|0\rangle \\
\quad &= \sqrt{1/2}\big( [c_{1,i}^\dagger]^2 - [d_{1,i}^\dagger]^2 + [c_{2,i}^\dagger]^2 - [d_{2,i}^\dagger]^2\big)|0\rangle
\label{eq:errortermcd}
\end{align}
in the heralded state \eqref{eq:dtmsvheraldedstate}.  In certain entanglement distribution architectures such as ZALM, this is mitigated by heralding at the receivers---the erroneous component of the heralded state provides vacuum to one of the receivers which can be filtered by the receiver's heralding process, e.g. by attempting to load the state into a heralded quantum memory.  Alternatively, we now show that one can directly generate a perfect heralded Bell pair with probability 1/8 by performing an entanglement swap between two anti-correlated pairs of the form \eqref{eq:noonstate}.  

\begin{figure}[h!]
\begin{quantikz}[row sep=0.4cm, column sep=0.3cm]
\setwiretype{c}\lstick[1]{$a_1^\dagger b_1^\dagger |0\rangle$} &
\ctrl[]{2} &
&
&
&
&
& 
\rstick[2]{$
|\Psi^\pm\rangle$} \\
\setwiretype{c}\lstick[1]{$a_2^\dagger b_2^\dagger |0\rangle$} &
&
\ctrl[]{2} &
&
&
& 
&
\\
\setwiretype{c}\lstick[1]{$|0\rangle$} &\control{} & & \ctrl{1}\gategroup[4,steps=5,style={dashed,inner ysep=2pt,inner xsep=3pt,yshift=-2pt}, label style={xshift=-0.75cm,yshift=-2cm}]{BSM} & \push{\scriptstyle\otimes}
\wire[d][2]{q} &
\setwiretype{q} & \wire[r][1]["c"{below,pos=0.5}, draw=none]{q} & 
\meterD[style={fill=black}]{}
\\[-0.25cm]
\setwiretype{c}\lstick[1]{$|0\rangle$} & &\control{} & \control{} & & \push{\scriptstyle\otimes}
\wire[d][2]{q} & \setwiretype{q}\wire[r][1]["c"{below,pos=0.5}, draw=none]{q} & 
\meterD{}
\\[-0.25cm]
\setwiretype{n} & & & &\push{\scriptstyle\otimes} & \setwiretype{q}  & 
\wire[r][1]["d"{below,pos=0.5}, draw=none]{q} & \meterD{}
\\[-0.25cm]
\setwiretype{n} & & & & & \push{\scriptstyle\otimes} & \setwiretype{q} \wire[r][1]["d"{below,pos=0.5}, draw=none]{q} &
\meterD[style={fill=black}]{}  
\end{quantikz}
\captionsetup{font=footnotesize,labelfont=footnotesize,justification=raggedright}
\caption{Entanglement swap of anti-correlated pairs to produce a Bell pair with success probability 1/8 using linear optics.}
\label{fig:noonswap}
\end{figure}
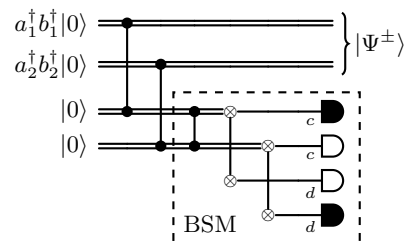

The method is shown in Fig. \ref{fig:noonswap}.  First, each anti-correlated pair \eqref{eq:noonstate} is combined with vacuum using a 50:50 beam-splitter to obtain two partial Bell states as expressed on the right side of \eqref{eq:dtmsvheraldedstate}.  We then perform an entanglement swap between the two sources using one of the outputs from each beam-splitter.  This is done by performing a partial-BSM in the $\{c,d\}$ basis, as shown in Fig. \ref{fig:noonswap}.  Since the partial-BSM only succeeds if opposite polarizations are detected, the error term $|\varepsilon^+\rangle$ is filtered out by the BSM, yielding a heralded Bell pair in the other two output modes (note that filtering of the error term requires that the BSM is performed in the diagonal basis, \emph{cf.} \eqref{eq:errortermab}-\eqref{eq:errortermcd}).  The partial-BSM succeeds with probability 1/2 given that a Bell state is provided by both sources, each of which produce a Bell state with probability 1/2; hence, the maximum success probability of the Bell state projection is 1/8.

The reader might note the resemblance of this operation to operations in fusion-based quantum computing (FBQC).  Indeed, since the input state $a_1^\dagger b_1^\dagger a_2^\dagger b_2^\dagger|0\rangle$ is simply 4 single-photon states with zero initial entanglement, this operation can be understood as part of a class of FBQC techniques for generating resource states \cite{BARTOLUCCI2021,PANKOVICH2024,FORBES2025}.  The method presented here is particularly well-suited to the application of entanglement distribution, as it does not require any active mode-conversion and the only feed-forward operations required are phase shifts that can be applied after distributing the Bell pair.

\section{Rate and fidelity of double-heralded multimode TMSV with realistic detectors}\label{sec:realistic}

\subsection{Double-heralded $M$-TMSV with realistic detectors}

In order to ensure high-fidelity heralded pairs with imperfect detectors, the mean photon number per mode $\mu$ must be tuned to sufficiently reduce errors from both higher order pairs and excess noise.  In this section, we characterize this optimization by evaluating the probability $\bar P_M^{(k,l)}$ of a double-heralding event and the fidelity of the projection of the heralded state onto the desired anti-correlated pair state in the idler modes.  The probability of one of the $M^2$ possible heralding events is
\begin{equation}\label{eq:heraldingprobabilitydef}
\bar P_{M}^{(k,l)} = \tr(E_{ab}^{(k,l)} |\psi\rangle\langle \psi |^{\otimes M}),
\end{equation}
where $E_{ab}^{(k,l)}$ represents a POVM element associated to a simultaneous single-click in detectors $k$ and $l$ on the signal modes of the $M$-TMSV and $M$-$\overline{\textup{TMSV}}$ arrays, respectively, with no other detections (\emph{cf.} Fig. \ref{fig:doubleheraldedmultimode}).  Without loss of generality, we can take $k=l=1$ and restrict our calculation to
\begin{equation}\label{eq:povmab}
E_{ab}^{(1,1)}=E^{(1)}_a\otimes E^{(1)}_b\otimes\big(E_a^{(0)}\otimes E_b^{(0)}\big)^{\otimes M-1}
\end{equation}
with subscript denoting the signal modes $a$ and $b$ of the TMSV and $\overline{\textup{TMSV}}$ sources, respectively, and the single superscripts on the right hand side denote the number of photons registered by the detector (in contrast to the double-superscripts on the left which denote the indices of the two detectors which registered a click).  For notational convenience, we have enumerated the sources in pairs $|\psi\rangle=|\textup{TMSV}\rangle|\overline{\textup{TMSV}}\rangle$ in \eqref{eq:heraldingprobabilitydef} and \eqref{eq:povmab}.  

It is assumed that the measurement realizing the POVM on the heralding modes does not interact with the idler modes, so that the post-measurement state can be obtained by taking the partial trace over the signal modes $\hat\rho_i=\tr_{s}(E_{ab}^{(k,l)}|\psi\rangle\langle \psi|^{\otimes M} )$ of the product Hilbert space $\mathcal H=\mathcal H_s\otimes \mathcal H_i$ of the system.  The heralded anti-correlated pair fidelity is then given by
\begin{equation}
F = \tr(|\Psi_{kl}\rangle\langle\Psi_{kl}|\hat\rho_i)=\frac{P_M^{(k,l)}}{\bar P_M^{(k,l)}}
\end{equation}
where
\begin{equation}\label{eq:trueheraldingprobabilitydef}
P_M^{(k,l)} = \tr \Big(\Big[ E_{ab}^{(k,l)} \otimes |\Psi_{kl}\rangle \langle \Psi_{kl}|\Big] |\psi\rangle\langle \psi |^{\otimes M}\Big)
\end{equation}
is the true pair generation probability, i.e. the probability of a successful heralding event followed by an ideal measurement of the desired anti-correlated pair in the idler modes.  Here, the projector $E_{ab}^{(k,l)}$ acts only on the subspace $\mathcal H_s$ spanned by the signal modes, and $|\Psi_{kl}\rangle\langle \Psi_{kl}|$ acts only on the idler subspace $\mathcal H_i$.

In this section, we are interested in understanding the performance of the source with realistic heralding detectors; hence, we model detection in each heralding mode $x=a,b$ via a POVM element of the general form
\begin{equation}\label{eq:povm1}
E^{(m)}_{x} = \sum_{n=0}^\infty p(m|n)|n\rangle\langle n|_{x}
\end{equation}
where $p(m|n)$ is the probability of registering $m$ counts in the detection mode given the presence of an $n$-photon Fock state.  Detection in each mode is modeled with quasi-PNR as an array of $K$ ideal single-photon detectors (SPDs) with efficiency $\eta_d$ and dark count probability $p_d$.  Following the derivation given in Appendix \ref{app:povm}, the probabilities of a null-count or single-count using this quasi-PNR detector model take the form
\begin{align}\label{eq:zeroclick}
p(0|n) &=  (1-\delta_0)(1-\eta_d)^n \\
\begin{split}\label{eq:singleclick}
p(1|n) &= \delta_1(1-\eta_d)^n \\
&\;\;\;\;+ \frac{1-\delta_2}{1-\alpha}\big[(1-\alpha\eta_d)^n - (1-\eta_d)^n\big]
\end{split}
\end{align}
where $\delta_0=1-p(0|0)$ is the probability that at least one click is recorded given a vacuum state in the detection mode, $\delta_1=p(1|0)$ represents the probability that exactly one count is recorded given that no photons in the signal mode are registered by the detector, $\delta_2$ represents the probability that at least one false secondary count is recorded given exactly one photon in the signal mode is registered by the detector, and the PNR availability $0\leq\alpha\leq 1$ represents the reduction in the availability of the detector to register a second photon count concurrent with another detection.  These quantities are derived rigorously based on the aforementioned detector array model in Appendix \ref{app:povm}, and take the values 
\begin{align}
\alpha&=(K-1)/K, \\
\delta_0 &= 1-(1 - p_d)^K \\
\delta_1 &= Kp_d(1-p_d)^{K-1},\\
\delta_2 &= 1 - (1-p_d)^{K-1}.
\end{align}
Note that within this model $\delta_1$ and $\delta_2$ are determined by the dark count probability $\delta_0$ and PNR availability $\alpha$.  That is, we have $\delta_1=[(1-\delta_0)^\alpha - (1-\delta_0)]/(1-\alpha)$ and $\delta_2=1 - (1-\delta_0)^\alpha $, reducing to $\delta_1\simeq\delta_0$ and $\delta_2\simeq \alpha\delta_0$ for $\delta_0\ll 1$.  Another quirk of modeling PNR as an array of SPDs with individual dark count probability $p_d$ is that the aggregate single dark count probability $\delta_1$ is constrained by PNR, with upper bound $\delta_1\leq \alpha^{\alpha/(1-\alpha)}$ decreasing to $1/e$ as $\alpha\to 1$, achieving the upper bound at $p_d=1/K$. 
  
For more general types of photon-counting detectors, the 4 parameters $\alpha,\delta_0,\delta_1,\delta_2$ above can be generalized and decoupled as independent fitting parameters to describe the behavior of a detector with partial PNR as in \cite{DAVIS2022}, though of course there is no guarantee that an arbitrary PNR detector will conform to the single-click and zero-click probabilities expressed by \eqref{eq:singleclick} and \eqref{eq:zeroclick} for all $n$.  The single-click probability for an ideal PNR detector is obtained in the limit $\alpha\to 1$ as
\begin{equation}\label{eq:singleclickpnr}
p_{\textup{PNR}}(1|n) = \delta_1 (1-\eta_d)^n + n(1-\delta_2) \eta_d(1-\eta_d)^{n-1}.
\end{equation}

\subsection{Pair probability and fidelity of double-heralded $M$-TMSV with realistic detectors}

To calculate the true pair generation probability $P_M^{(1,1)}$, we can replace the projector $|\Psi_{11}\rangle\langle \Psi_{11}|$ with the equivalent projection applied before the unitary $U_M$, namely $|1,1\rangle\langle 1,1|_{11}$.  The projection of the $M$-fold tensor product of \eqref{eq:dtmsv} becomes
\begin{equation}\label{eq:noonprojection}
|1,1\rangle\langle 1,1|_{11}|\psi\rangle\langle \psi|^M = 
\frac{\tanh^4(r)}{\cosh^{4M}(r)} |1,1;1,1\rangle\langle 1,1;1,1|_{11}
\end{equation}
where the semi-colon separates the signal and idler modes of the heralded modes $k=l=1$.
The trace \eqref{eq:trueheraldingprobabilitydef} is then
\begin{align}
\begin{split}
P_M^{(1,1)} &= (1-\lambda)^{2M}\lambda^2 p(1|1)^2 p(0|0)^{2M-2} \\
&= \frac{\mu^2}{(1+\mu)^{2+2M}}p(1|1)^2 p(0|0)^{2M-2}.
\end{split}
\end{align}
Substituting the click probabilities yields
\begin{equation}\label{eq:truedoubleheraldingprobabilityeq}
P_M^{(1,1)} = \frac{(\eta\mu)^2(1-\delta_0)^{2M-2}}{(1+\mu)^{2+2M}}\left(1 - \delta_2 + \delta_1\frac{1-\eta}{\eta}\right)^2
\end{equation}
where $\eta$ is the overall efficiency from pair production to detection.

The raw heralding probability $\bar P_M^{(1,1)}$ can be calculated from \eqref{eq:tmsv2}, \eqref{eq:heraldingprobabilitydef}, and \eqref{eq:povm1} and takes the form of a double summation over single-click probabilities which factors as
\begin{equation}
\bar P^{(1,1)}_M = \big(\bar P^{(1)}_M\big)^2
\end{equation}
where
\begin{equation}
\bar P^{(1)}_M = (1-\lambda)^{M}\sum_{k=0}^\infty \lambda^k p(1|k) \bigg(\sum_{j=0}^\infty \lambda^j p(0|j)\bigg)^{M-1}
\end{equation}
is the probability of registering a single count in a specified signal mode of an $M$-TMSV state (and registering no counts in all other signal modes).

Substituting the click probabilities yields two geometric series which can be evaluated in closed form
\begin{align}\label{eq:singleheraldingprobabilityclosedform}
\bar P_M^{(1)} = \frac{\eta\mu(1-\delta_0)^{M-1}}{(1+\eta\mu)^M(1+\alpha\eta\mu)}\Big(1-\delta_2 + \delta_1\frac{1 + \alpha\eta\mu}{\eta\mu}\Big).
\end{align}
Thus, we obtain the fidelity of the heralded anti-correlated pair with realistic heralding detectors as
\begin{align}\label{eq:doubleheraldedfidelityeq}
\begin{split}
F &= \frac{(1+\eta\mu)^{2M}(1+\alpha\eta\mu)^2}{(1+\mu)^{2+2M}} \\
&\qquad\qquad \times \frac{[\delta_1(1-\eta)\mu + (1-\delta_2)\eta\mu]^2}{[\delta_1(1+\alpha\eta\mu) + (1-\delta_2)\eta\mu]^2}.
\end{split}
\end{align}

Eqs. \eqref{eq:truedoubleheraldingprobabilityeq} and \eqref{eq:doubleheraldedfidelityeq} represent the main results of the analysis of double-heralded $M$-TMSV sources without multiplexing.  To facilitate later calculations of the multiplexed pair generation probability, we note that since $\bar P_M^{(1)}$ represents only 1 of $M$ possible modes in which we can herald a single pair from an $M$-TMSV source, the total single-photon herald probability is (\emph{cf.} \eqref{eq:singleheraldprobability})
\begin{equation}
p_M = M\bar P_M^{(1)}.
\end{equation}

\subsection{Fidelity and mean photon number}\label{sec:limitingcases}
In this section we use the results above to develop expressions for the mean photon number per mode needed to achieve a desired state fidelity.  To this end, it is useful to consider some limiting cases.  First, note that we have expressed the results \eqref{eq:truedoubleheraldingprobabilityeq} and \eqref{eq:doubleheraldedfidelityeq} such that the trailing factors approach unity in the absence of excess noise, yielding simplified expressions which illustrate the importance of driving both the detection efficiency $\eta$ and PNR availability $\alpha$ closer to unity to allow larger $\mu$.

In the limit of ideal PNR $(\alpha\to 1)$ and no excess noise $(\delta_0=\delta_1=\delta_2=0)$ the fidelity becomes
\begin{equation}\label{eq:fidelitypnr}
F_{\textup{PNR}} = \left(\frac{1 + \eta\mu}{1+\mu}\right)^{2+2M},
\end{equation}
quantifying the minimal requirement on detection efficiency $\eta$ as the mean photon number increases. Explicitly, the general expression for the maximal mean photon number at a given fidelity reduces to
\begin{equation}\label{eq:optimalmupnr}
\hat\mu_{\textup{PNR}}(\eta;F) = \frac{1 - F^{1/(2+2M)}}{F^{1/(2+2M)} - \eta}
\end{equation}
provided $\eta^{2+2M}<F<1$.  If $\eta^{2+2M} \geq F$ than the fidelity is strictly greater than $F$ for all $\mu$, and one can maximize the true pair generation rate \eqref{eq:truedoubleheraldingprobabilityeq} by taking $\mu = 1/M$ just as in \eqref{eq:optimalmumtmsv}.

For non-ideal PNR, one can find the maximum allowed mean photon number subject to a given fidelity requirement by rewriting \eqref{eq:doubleheraldedfidelityeq} without excess noise as
\begin{equation}
(1+\mu)^{M+1}F^{1/2} = (1+\eta\mu)^M(1+\alpha\eta\mu),
\end{equation}
yielding a polynomial of degree $M+1$ in $\mu$.  In general, we must resort to numerical methods to find the largest positive root, clamping the result to the upper bound $\mu \leq 1/M$ to maximize single-pair generation.  

Note that the reduction in fidelity from non-ideal PNR availability $\alpha<1$ in \eqref{eq:doubleheraldedfidelityeq} is independent of $M$.  Taking the limit $\eta\to 1$ we obtain a general upper bound on the fidelity based solely on the PNR availability
\begin{equation}
F\leq \Big(\frac{1+\alpha\mu}{1+\mu}\Big)^2.
\end{equation}
This leads to an upper bound on the mean photon number per mode allowed at a given target fidelity
\begin{equation}\label{eq:pnrmubound}
\mu(F) \leq \frac{1 - F^{1/2}}{F^{1/2} - \alpha},
\end{equation}
that holds when $\alpha < F^{1/2}$.  This constrains the pair generation rate based on the number of detectors $K=1/(1-\alpha)$ in a quasi-PNR detection scheme (\emph{cf.} Fig. \ref{fig:quasipnrmultiplexing}).

\subsection{Maximum pair fidelity with excess noise}\label{sec:excessnoise}
In the absence of excess noise, one can improve the pair fidelity arbitrarily close to unity by lowering the mean photon number per mode $\mu$ to reduce noise from higher order pairs.  This is useful for applications requiring an exceptionally low second-order correlation $g^{(2)}(0)$.  The presence of excess noise puts a limit on the maximum fidelity one can achieve in this manner; for sufficiently small $\mu$ the fidelity becomes limited by the fraction of events triggered by excess noise.

\begin{figure}[h!]
\centering\includegraphics[width=8.5cm]{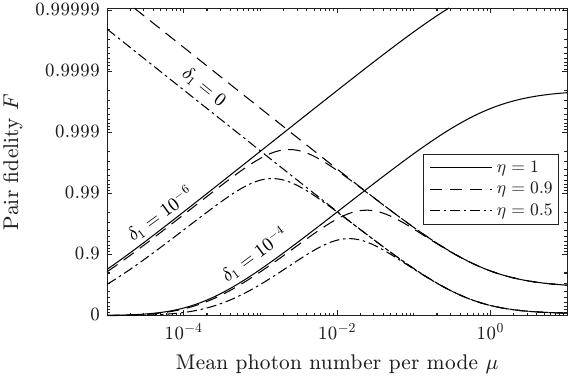}
\captionsetup{font=footnotesize,labelfont=footnotesize,justification=raggedright}
\caption{Fidelity as a function of mean photon number with single noise count probability $\delta_1=10^{-4}$ and $10^{-6}$, shown in two groups from bottom to top, following the asymptote at $\delta_1=0$ going off to the top left.  Three curves at each noise level are shown, with detection efficiency $\eta=0.5,0.9$, and $1$ as indicated in the legend.}
\label{fig:pnrfidelitynoise}
\end{figure}

Fig. \ref{fig:pnrfidelitynoise} shows the dependence of the fidelity of double-heralded TMSV $(M=1)$ on the mean photon number per mode for several noise count probabilities $\delta_1$. The asymptotes towards the upper left shows the unbounded fidelity with $\delta_1=0$ for $\eta=0.5$ and $\eta=0.9$.  The horizontal asymptotes on the lower right reflect the lower limit $F\to \eta^{2+2M}$ as $\mu\to \infty$.  Note that for $\delta_1>0$, the fidelity is virtually unaffected by the noise as $\mu$ decreases nearly until reaching the minimum $\mu$ before the fidelity starts to decline.  This global maximum fidelity increases with detection efficiency; however, the increase does have a finite bound in the limit $\eta = 1$ as seen in the horizontal asymptotes at the top right of the figure, which can be calculated as
\begin{equation}
F \leq \left(\frac{1-\delta_2}{\delta_1 + 1-\delta_2}\right)^2\sim 1-2\delta_1.
\end{equation}

In the presence of excess noise, the minimum mean photon number per mode before the fidelity starts to decline can be calculated by differentiating \eqref{eq:doubleheraldedfidelityeq}.  This yields an expression for the minimal $\mu$ based on the noise count probability provided in Appendix \ref{app:optimalmu}. Fig. \ref{fig:maxfidelitynoise} shows the dependence of the maximum fidelity on the noise count probability and detection efficiency for double-heralded $M$-TMSV with $M\leq 4$.  

\begin{figure}[h!]
\centering\includegraphics[width=8cm]{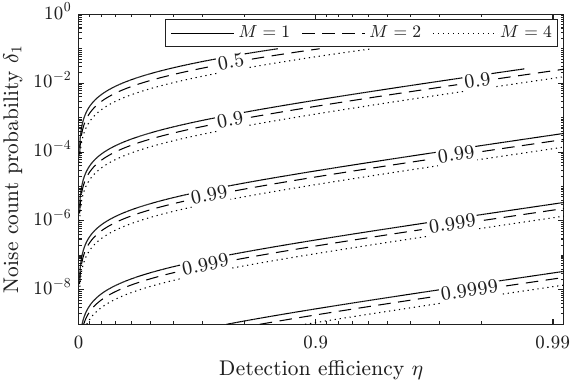}
\captionsetup{font=footnotesize,labelfont=footnotesize,justification=raggedright}
\caption{Contours show the maximum achievable fidelity based on the noise count probability $\delta_1$ and detection efficiency $\eta$.}
\label{fig:maxfidelitynoise}
\end{figure}

Note that the reduction in fidelity due to excess noise coming from the second factor in \eqref{eq:doubleheraldedfidelityeq} is independent of $M$; however, since larger $M$ require a smaller mean photon number per mode to achieve the same pair fidelity, \emph{cf.} \eqref{eq:fidelitypnr}, the effects of excess noise put stricter upper bounds on fidelity with increasing $M$.  This leads to an increased requirement on the detection efficiency needed to achieve the target fidelity with a given excess noise probability.

\subsection{Multiplexed pair probability for double-heralded TMSV with detection losses}\label{sec:multiplexing}

\begin{figure}[h!]
\centering\includegraphics[width=8.5cm]{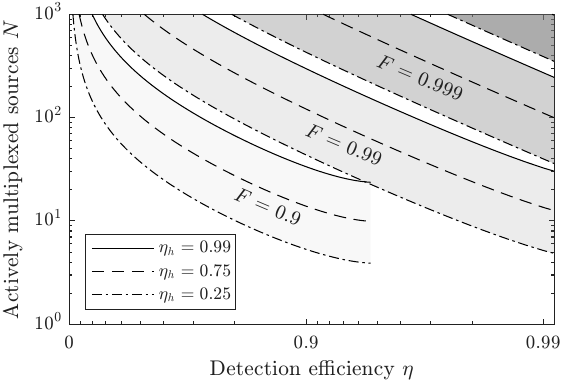}
\captionsetup{font=footnotesize,labelfont=footnotesize,justification=raggedright}
\caption{Number of actively multiplexed TMSV sources needed to achieve quasi-deterministic operation as a function of detection efficiency.  Each shaded band corresponds to a single target pair fidelity.  The curves for $F=0.9$ are truncated once reaching their minima at the maximum pair generation rate $\mu = 1$.  The power law relation at high fidelity is described by $N\propto (1-\eta)$.}
\label{fig:multiplexedpairgeneration}
\end{figure}

Finally, we return to the question of determining the number of multiplexed sources required for quasi-deterministic operation.  With the inclusion of detection losses, the mean photon number per mode must be correspondingly reduced to achieve the target fidelity, increasing the number of multiplexed sources needed to compensate.  Fig. \ref{fig:multiplexedpairgeneration} shows the relation between detection efficiency and the required multiplexing factor required to achieve operation with double-heralding probability $\eta_h = 0.25, 0.8$ and $0.99$, assuming PNR detection ($\alpha=1$) and a maximally-connected multiplexing configuration \eqref{eq:maximalmultiplexingrequirement} with $M=1$.

Examination of Fig. \ref{fig:multiplexedpairgeneration} yields several useful observations.  Note that one can generally reduce the infidelity by a factor of $\sim$10 by reducing the generation probability from 99\% (essentially fully deterministic) to about 25\%.  For example, with $90\%$ detection efficiency and 44 multiplexed sources one can produce anti-correlated pairs with a pair generation probability $>99\%$ ($25\%$) at $90\%$ ($99\%$) fidelity.  For $F\gtrsim 0.99$, we observe that the required multiplexing scales in direct proportion to the loss $(1-\eta)$, demonstrating how critical the heralding efficiency is to enabling a high-fidelity, quasi-deterministic source.  Regarding the example above, increasing the detection efficiency from $90\%$ to $95\%$ targeting a $25\%$ generation probability at $99\%$ fidelity reduces the required number of multiplexed sources from 44 to just 22.  Finally, we observe that the required multiplexing factor approaches the lossless limit roughly as the detection efficiency reaches the target fidelity $\eta \sim F$.

\begin{figure}[h!]
\centering\includegraphics[width=8.5cm]{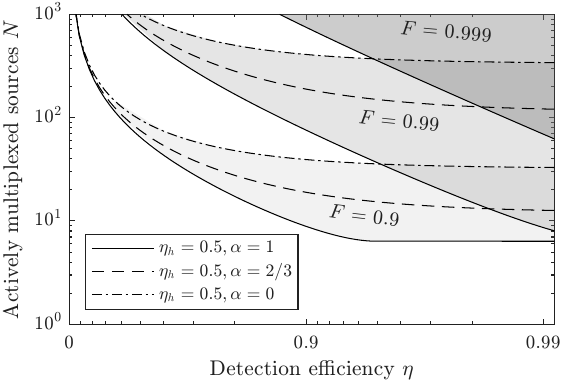}
\captionsetup{font=footnotesize,labelfont=footnotesize,justification=raggedright}
\caption{Active multiplexing factor $N$ needed to achieve double-heralding probability $\eta_h=0.5$ with quasi-PNR ranging from $\alpha=1$ (PNR) to $\alpha = 0$ (no PNR).}
\label{fig:quasipnrmultiplexing}
\end{figure}

However, for detection with quasi-PNR modeled as an array of $K$ SPDs, the gain one can achieve through increased detection efficiency is constrained by the PNR availability $\alpha=1-1/K$.  Fig. \ref{fig:quasipnrmultiplexing} shows the effect of quasi-PNR with $\alpha < 1$ on the required multiplexing factor needed to achieve $50\%$ generation probability.  Observe that with $\alpha<1$, the required multiplexing factor does not continue to decrease with the detection loss $(1-\eta)$, instead saturating at a minimum multiplexing factor which depends on $\alpha$ (this minimum can be calculated using \eqref{eq:pnrmubound}).  For example, with 90\% detection efficiency one can achieve a 50\% generation probability at 99\% fidelity using 75 TMSV sources with PNR.  With $\alpha = 2/3$, corresponding to an array of 3 SPDs per mode, one cannot achieve this rate and fidelity with less than 114 sources, and with $\alpha = 0$ (no PNR) the requirement increases to no fewer than 335 sources.

\begin{figure}[h!]
\centering\includegraphics[width=8.5cm]{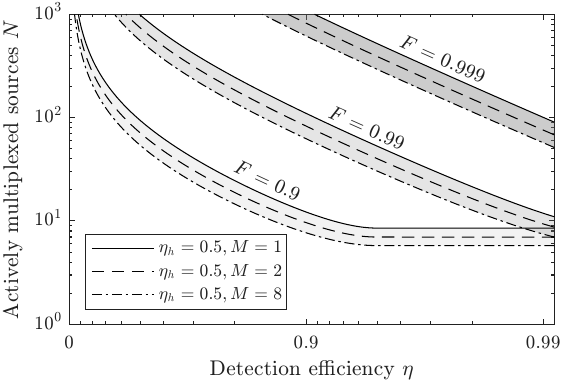}
\captionsetup{font=footnotesize,labelfont=footnotesize,justification=raggedright}
\caption{Actively multiplexed $M$-TMSV sources $N$ needed to achieve double-heralding probability $\eta_h=0.5$ in a bipartite multiplexing configuration using multimode TMSV with $M\leq 8$.}
\label{fig:bipartitemultiplexing}
\end{figure}

Finally, Fig. \ref{fig:bipartitemultiplexing} shows the number of actively multiplexed $M$-TMSV sources needed in a bipartite multiplexing configuration for double-heralded $M$-TMSV with $M\leq 8$. Recall that in the bipartite ZALM configuration, the number of spectral modes $n=N/2$ is half of the active multiplexing factor.  For example, we find that with 90\% detection efficiency and $M=2$, a 50\% double-heralding probability targeting 99\% pair fidelity can be achieved with $N/2=41$ spectral modes.  Interestingly, this is in precise agreement with the recent results of Shapiro \emph{et. al.} for the swap-heralded D-TMSV source with cross-island heralding---based on a 99\% Bell state fidelity after 20 dB transmission losses---despite the fact that we use the anti-correlated pair fidelity without transmission losses as our target metric.  The use of multimode TMSV with $M=2$ reduces the active multiplexing requirement by a factor of about 0.77 compared to $M=1$, and again we find a similar factor as we increase from $M=2$ to $M=8$.  Thus, as in the ideal case $\eta=1$ summarized in Fig. \ref{fig:multiplexedefficiency}, we find that $M=1$ yields the most efficient use of resources in terms of minimizing the total multiplexing factor $MN$ needed to achieve a specified double-heralding rate and fidelity.

\section{Conclusion}

A double-heralding technique was presented for producing heralded Bell pairs from spontaneous parametric down-conversion based on the observation that the partial Bell pairs produced by previously proposed cascaded SPDC and ZALM schemes are best understood simply as heralded anti-correlated pairs.  The double-heralding scheme presented in Sec. \ref{sec:doubleheralded}, as well as its generalization to multimode $M$-TMSV arrays in Sec. \ref{sec:doubleheraldedmtmsv}, reduce the mode-sorting optics needed on the heralding path which can increase the overall detection efficiency $\eta$.  The number of multiplexed sources needed to achieve a specified double-heralding rate and fidelity was found to scale with the detection loss $1-\eta$ (Sec. \ref{sec:multiplexing}), demonstrating the importance of minimizing losses on the heralding path. It was also shown that previous swap-heralded ZALM and cascaded SPDC schemes are otherwise equivalent to double-heralded $M$-TMSV with $M=2$.  The analysis found that although using $M$-TMSV with $M>1$ offers a reduction in the number of modes which must be \emph{actively} multiplexed with mode-converters, double-heralding with individual TMSV sources yields the most resource-efficient implementation in terms of minimizing the total number of source modes and detectors required to achieve a specified rate and fidelity.

As a corollary, the results show that for ZALM implementations which do not allow any spatial mode switching, $M$-TMSV with $M>1$ reduces the number of actively switched spectral modes $N_I$ required to achieve a specified rate and fidelity (Fig. \ref{fig:bipartitemultiplexing}), at the cost of increasing the total number of source modes and detectors by a factor of $M$.  Alternatively, since $M$-TMSV with $M=1$ yields the most efficient implementation in terms of source modes and detectors, one can outperform a single 2-TMSV ZALM source by actively multiplexing the outputs of 2 identical 1-TMSV ZALM sources, with both configurations using an identical total number of source modes, detectors, and spectral islands.  Of course, as this requires spatially multiplexing two ZALM sources, it should perhaps be understood as single-added-loss multiplexing (SALM), with the conclusion that 1-TMSV SALM can outperform 2-TMSV ZALM given the same number of sources, detectors, and spectral islands.  In practice, optimizing the source design ultimately comes down to an engineering problem which considers implementation-based constraints and trade-offs between the different configurations outlined in this paper.  For example, given device-specific performance parameters (component losses, noise, PNR, etc.), the analytical results obtained in this paper then provide a framework to determine which implementations will yield the best heralded pair rate and fidelity.

The analysis in this paper is based on a study of the anti-correlated pair state produced by the source alone, examining the state of the heralded idler modes without any additional losses or channel effects.  Nevertheless, our results agree very closely with the analysis of the ZALM source presented by Shapiro \emph{et. al.} based on an analysis of the Bell state fidelity including transmission losses.  A complete analysis of the overall performance of entanglement distribution with ZALM based on double-heralded TMSV, including an analysis of the quantum state loaded into memory after receiver losses, is left for future consideration.

\section*{Acknowledgements}

The authors would like to thank Clark Embleton and Jeff Shapiro for their interest in this work, and for several insightful discussions relating this work to their ZALM cross-island heralding scheme.  The authors also thank Jason Saied for a number of discussions related to early versions of this paper.  This research was supported by the NASA Glenn Research Center (GRC) Center Innovation Fund (CIF),  NASA Convergent Aeronautics Solutions QUARC project, the NASA Space Technology Graduate Research Opportunities (NSTGRO) Program, and the GRC Communications \& Intelligent Systems Division.
\vfill

\bibliography{ChahineBibliography2022}

\appendix
\section{Symmetries of $N$-fold TMSV}\label{app:symmetry}

\begin{figure}[h!]
\begin{quantikz}[row sep=0.25cm, column sep=0.25cm]
&\gate[4]{U_N} &
\gate{\textup{TMSV}} &&&
\\
&&
\gate{\textup{TMSV}} &&&
\\  
\setwiretype{n}& & \vdots & &
\\
&&
\gate{\textup{TMSV}} &&&
\end{quantikz}=\begin{quantikz}[row sep=0.25cm, column sep=0.25cm]
&&&
\gate{\textup{TMSV}} &
\gate[4]{U_N} &
\\
&&&
\gate{\textup{TMSV}} &
&
\\  
\setwiretype{n}&&& \vdots & &
\\
&&&
\gate{\textup{TMSV}} & &
\end{quantikz}
\captionsetup{font=footnotesize,labelfont=footnotesize,justification=raggedright}
\caption{Equivalence of unitaries applied to signal modes and idler modes of $N$-fold tensor products of TMSV states.}
\label{fig:unitarytmsvequivalence}
\end{figure}
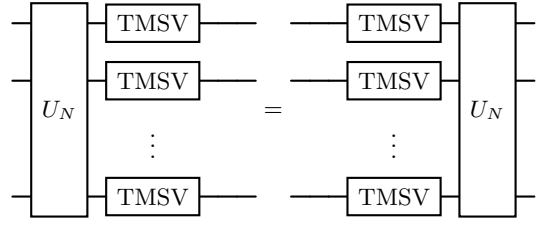

In this appendix, we prove that applying a unitary $U_N$ to either the signal modes or idler modes of an $N$-fold tensor product of TMSV states yields completely equivalent states as expressed in the output modes of the system (Fig. \ref{fig:unitarytmsvequivalence}).  

To prove this, we show that the Hamiltonian associated to the $N$-fold tensor product
\begin{equation}\label{eq:nfoldtmsv}
H_{\textup{int}}^{(N)} = \sum_{j=1}^{N} a_{j,s}b_{j,i} + \textup{\emph{h.c.}}
\end{equation}
is invariant under the action of the unitary $[U_N]_s \otimes [U_N^{-1}]_i$, \emph{i.e.}, the unitary obtained by independently applying the $N$-dimensional unitary $U_N$ to the signal modes and its inverse $U_N^{-1}$ to the idler modes.  The result then follows from the identity
\begin{align}
[U_N]_s H_{\textup{int}}^{(N)} &= [U_N]_i[U_N^{-1}]_i[U_N]_s H_{\textup{int}}^{(N)} \\
&= [U_N]_i H_{\textup{int}}^{(N)}.
\end{align}
The proof is a simple calculation, letting $u^m_{\;\;n}$ and $v^m_{\;\;n}$ denote the matrix elements of $U_N$ and $U_N^{-1}$, respectively,
\begin{align}
[U_N^{-1}]_i[U_N]_s H_{\textup{int}}^{(N)} &= \sum_{j=1}^N \sum_{k,\ell=1}^N \big(u^k_{\;\,j} \tilde a_{k,s}\big)\big(v^\ell_{\;\,j} \tilde b_{\ell,i}\big) \\
&= \sum_{k,\ell=1}^N \sum_{j=1}^N u^k_{\;\,j}\bar u^j_{\;\,\ell} \tilde a_{k,s}\tilde b_{\ell,i} \\
&= \sum_{k=1}^N \tilde a_{k,s}\tilde b_{\ell,i}
\end{align}
which is identical to the state expressed in the input modes \eqref{eq:nfoldtmsv} (we have left off the hermitian conjugate terms for brevity).  Explicitly, the second line uses the identity $v^\ell_{\;\,j}=\bar u^j_{\;\,\ell}$ following from $U_N^{-1}=U_N^\dagger$ and the second line again uses unitarity to observe that $\sum u^k_{\;\,j}\bar u^j_{\;\,\ell} = \delta^k_{\;\ell}$ is the Kronecker delta.  Note that the actual mode labels $a,b$ are arbitrary, and the proof applies whether or not the sum in \eqref{eq:nfoldtmsv} includes any combination of terms with $a_{j,s}b_{j,i}, b_{j,s}a_{j,i}, a_{j,s}a_{j,i}$, etc., provided all of the annihilation operators represent orthogonal modes and appear only once in the sum (as in \eqref{eq:dtmsvhamiltonian}).

\section{Derivation of detection POVM}\label{app:povm}

The analysis in Sec. \ref{sec:realistic} is based on a model for PNR detection with an imperfect detector modeled as an array of $K$ ideal single-photon detectors (SPDs) fed from a single optical mode by a $1\times K$ unitary splitting operation $U_K$ preceded by a beam-splitter with transmissivity $\eta_d$ to model a uniform detection efficiency, assuming vacuum in all of the unused input modes.  

The general solution for the POVM associated to such an arrangement in the absence of dark counts was derived by Davis \emph{et al.} \cite{DAVIS2022}.  In this appendix, we first generalize this derivation in to include dark counts.  A somewhat subtle application of the binomial theorem then yields an expression for the single-click probabilities $p(1|n)$ via the geometric sum in \eqref{eq:singleclick} in place of sums involving binomial coefficients, the representation as a geometric sum being crucial to the analytical calculations in Sec. \ref{sec:realistic}.

Following \cite{DAVIS2022}, we consider the action of the unitary $U_K$ which splits a photon in the input mode into an equal superposition of the $K$ output modes which impinge on the ideal SPDs.  An $n$-photon Fock state in the input mode transforms via the multinomial theorem as
\begin{equation}
U_K|n\rangle =\frac{\sqrt{n!}}{\sqrt{K^n}}\sum_{k_1+...+k_K=n} \frac{1}{\sqrt{k_1!\cdots k_K!}} |k_1,...,k_K\rangle.
\end{equation}
Thus, the joint probability $P(k_1,...,k_K)$ of finding $k_i$ photons in the $i$-th output mode is non-vanishing only when $\sum k_i = n$ where it is given by a multinomial distribution
\begin{equation}
P(k_1,...,k_K) = \frac{1}{K^n}\frac{n!}{k_1!\cdots k_K!}, \qquad \sum k_i = n.
\end{equation}
It follows that the probability that exactly $m$ of the output modes receive at least one photon is given by
\begin{equation}
P_K(m|n) = \frac{m!}{K^n}{K \choose m}\stirling{n}{m}
\end{equation}
where the Stirling number of the second kind
\begin{equation}
\stirling{n}{m} = \sum_{j=1}^m \frac{(-1)^{m-j} j^n}{(m-j)!j!}
\end{equation}
counts the number of ways to partition a set of $n$ labeled objects into $m$ non-empty (unlabeled) sets, so that
\begin{equation}
m!{K \choose m}\stirling{n}{m}
\end{equation}
counts the number of unique ways of sorting $n$ labeled objects into $K$ labeled bins such that precisely $m$ bins are non-empty.

In the presence of losses, the probability $P_K(m|n;\eta_d)$ of an $m$-fold coincidence given an $n$-photon input Fock state takes the form
\begin{equation}
P_K(m|n;\eta_d) = \sum_{j=0}^n P^m_{K,j} L^j_{\;n}
\end{equation}
where the loss matrix $L^j_{\;n}$ is given by
\begin{equation}
L^j_{\;n}={n \choose j} \eta_d^j(1-\eta_d)^{n-j}
\end{equation}
and the coefficient matrix $P^m_{K,j}=P_K(m|j)$ represents the POVM in the absence of losses.  If we further assume that each SPD can trigger a dark count with probability $p_d$, the probability $P(m|n;\eta_d,p_d)$ of an $m$-fold coincidence given an $n$-photon input Fock state becomes
\begin{equation}
P_K(m|n;\eta_d,p_d)=\sum_{k=0}^m\sum_{j=0}^n D^m_{K,k} P^k_{K,j} L^j_{\;\, n}
\end{equation}
where the dark count matrix
\begin{equation}
D^m_{K,k} = {K-k \choose m-k} p_d^{m-k} (1-p_d)^{K-m}
\end{equation}
describes the probability of $m-k$ dark counts assuming exactly $k$ of the $K$ ideal SPDs simultaneously receive at least one photon.

In this work, we need only the zero-count and single-count coefficients.  The zero-count coefficients reduce to
\begin{equation}
p(0|n)=D_{K,0}^0 P_K(0|n;\eta_d)
\end{equation}
where $D_{K,0}^0=(1 - p_d)^K$ and $P_K(0|n;\eta_d)=(1-\eta_d)^n$.  The single-count probabilities reduce to
\begin{equation}
p(1|n) = D_{K,0}^1 P_K(0|n;\eta_d) + D_{K,1}^1P_K(1|n;\eta_d)
\end{equation}
with
\begin{align}
\begin{split}
P_K(1|n;\eta_d) &= \sum_{j=1}^n \frac{1}{K^{j-1}}{n \choose j}\eta_d^j(1-\eta_d)^{n-j} \\
&= K(1 - \frac{K-1}{K}\eta_d)^n - K(1-\eta_d)^n
\end{split}
\end{align}
where the second identity follows from rearranging terms and employing the binomial theorem.

Simplifying these expressions, we model the zero- and single-click probability of a detector with partial PNR and excess noise in the form 
\small
\begin{align}\label{eq:zeroclicka}
p(0|n) &= \nu_0(1-\eta_d)^n \\
\begin{split}
p(1|n) &= \delta(1-\eta_d)^n + \frac{\nu_2}{1-\alpha}\big[(1-\alpha\eta_d)^n - (1-\eta_d)^n\big]
\end{split}\label{eq:singleclicka}
\end{align}
\normalsize
where we have defined 
\begin{align}
\nu_0 &\equiv D_{K,0}^0= (1-p_d)^K, \\
\nu_2 &\equiv D^1_{K,1}=(1-p_d)^{K-1}, \\
\delta_1 &\equiv D^1_{K,0} = Kp_d(1-p_d)^{K-1}, \\
\alpha &\equiv (K-1)/K,
\end{align}
to obtain the result as expressed in Sec. \ref{sec:realistic}, substituting $\nu_0 = 1-\delta_0$ and $\nu_2=1-\delta_2$.  To model an ideal PNR detector, we can take the limit as $\alpha\to 1$ while keeping $\delta_1,\nu_2$ fixed to obtain
\begin{equation}
p_{\textup{PNR}}(1|n) = \delta_1(1-\eta_d)^n + n\nu_2 \eta_d(1-\eta_d)^{n-1}.
\end{equation}

\section{Minimum mean photon number with excess noise}\label{app:optimalmu}

As discussed in Sec. \ref{sec:excessnoise}, the presence of excess noise $\delta_1>0$ imposes a minimum mean photon number per mode $\mu_0$ below which any further reduction in $\mu$ will degrade the anti-correlated pair fidelity by introducing vacuum into the heralded state.  This minimum $\mu_0$ can be calculated by differentiating \eqref{eq:doubleheraldedfidelityeq} and finding the roots of the resulting polynomial equation.  Instead of differentiating \eqref{eq:doubleheraldedfidelityeq} directly, we instead extremize the function
\begin{equation}
f(\mu) = \frac{(1+\eta\mu)^M(1+\alpha\eta\mu)\mu}{(1+\mu)^{M+1}(1+\beta\eta\mu)}
\end{equation}
with $\beta = \alpha + (1-\delta_2)/\delta_1$.  This reduced function satisfies $f(\mu)\propto F^{1/2}$ provided $\delta_1>0$ and hence has the same local extrema.  The solutions $df/d\mu=0$ are given by the roots of the polynomial equation
\begin{equation}
(1+\eta\mu)^{M-1}(A\mu^3 + B\mu^2 + C\mu + 1) = 0,
\end{equation}
with
\begin{align}
\begin{split}
A &= \alpha\eta^2 - \beta\eta^2[1 - \alpha\eta + \alpha M(1-\eta)], \\
B &= \alpha\eta[1+2\eta-M(1-\eta)] - \beta\eta[1-\alpha\eta + M(1-\eta)], \\
C &= \eta(1+2\alpha) - M(1-\eta).
\end{split}
\end{align}
The first factor contributes only the negative multiple root $\mu=-1/\eta$, and so any positive roots must be roots of the cubic.  The result can be expressed using Cardano's formula
\begin{equation}
\mu_0 = (-v + \sqrt{\Delta})^{1/3} + (-v-\sqrt{\Delta})^{1/3} - \frac{B}{3A}
\end{equation}
with
\begin{align}
u &= \frac{C}{3A} - \frac{B}{3A^2} \\
v &= \frac{B^3}{27 A^3} - \frac{BC}{6A^2} + \frac{1}{2A} \\
\Delta &= u^3 + v^2
\end{align}
and we have chosen the only solution which yields a positive real root.  In certain edge cases with $\alpha,\eta\sim 1$ and large $\delta_1$, the discriminant $\Delta$ becomes negative, in which case there are no positive real roots (the fidelity has no positive local maximum in this case, and simply increases monotonically with $\mu$).

\end{document}